\documentclass{egpubl}
\usepackage{eurovis2026}

\SpecialIssuePaper         %

\CGFccby

\usepackage[T1]{fontenc}
\usepackage{dfadobe}  

\usepackage{cite}  %

\usepackage{enumitem}
\usepackage{booktabs}
\usepackage{amssymb}
\usepackage{xcolor}
\usepackage{caption}

\BibtexOrBiblatex
\electronicVersion
\PrintedOrElectronic
\ifpdf \usepackage[pdftex]{graphicx} \pdfcompresslevel=9
\else \usepackage[dvips]{graphicx} \fi
\graphicspath{{style/}}

\usepackage{egweblnk}
\usepackage{cleveref}
\crefalias{appendix}{section}
\crefname{appendix}{Appendix}{Appendices}
\Crefname{appendix}{Appendix}{Appendices}

\newif\ifshowmain
\showmaintrue
\newif\ifshowappendix
\showappendixfalse

\usepackage{enumitem}

\usepackage[group-separator={$,$}]{siunitx}
\usepackage{xspace}
\newcommand{\systemName}{HiFigAtlas\xspace}

\newcommand{\journalName}{\textit}

\newcommand{\taxon}{\textsl}

\newcommand{\percentageWithRawData}[2]{{#1}}

\makeatletter
\newcommand{\iflabelexists}[3]{%
  \@ifundefined{r@#1}{#3}{#2}%
}
\makeatother

\usepackage{multirow}

\newcommand{\nTaxaFinalAll}{\num{52}\xspace}

\newcommand{\nTaxaFinalAllVis}{\num{47}\xspace}
\newcommand{\nTaxaFinalLeafVis}{\num{40}\xspace}
\newcommand{\nTaxaFinalFirstLevelVis}{\num{20}\xspace}

\newcommand{\paperTitleShort}{How Historians Use Visualization}
\newcommand{\paperTitleLong}{How Historians Use Visualization: A Corpus‐Driven Mixed‐Methods Study}
\ifshowmain
\title[\paperTitleShort]%
    {\paperTitleLong}
\else
\title[\paperTitleShort]%
    {Appendix to\\ ``\paperTitleLong''}
\fi

\author[X. Chen et al.]
{
    \parbox{\textwidth}{
        \centering
        X.\,Chen$^{1,2}$\orcid{0009-0006-0700-4575},
        Y.\,Zhang$^{3,*}$\orcid{0000-0002-9035-0463},
        W.\,Zheng$^{4}$\orcid{0009-0000-4209-9094},
        C.\,Ma$^{1,2}$\orcid{0009-0001-8612-2492},
        and X.\,Yuan$^{1,2,*}$\orcid{0000-0002-7233-980X}
    }
    \\
    \parbox{\textwidth}{
        \centering
        $^1$National Key Laboratory of General Artificial Intelligence, School of Intelligence Science and Technology, Peking University, China\\
        $^2$PKU-WUHAN Institute for Artificial Intelligence, China\\
        $^3$Department of Computer Science, University of Oxford, UK\\
        $^4$School of Data Science, Fudan University, China\\
        $^*$corresponding authors
    }
}

\begin{document}

\maketitle
\ifshowmain
\begin{abstract}
Visualization in historical research is shifting from isolated attempts to systematic practices.
However, data-driven evidence about how historians actually use visualization remains scarce.
We present a corpus-driven, mixed-methods study that combines analysis of images from \num{4142} research articles across history and digital humanities journals with a collaboratively developed visualization taxonomy and a semi-automatic labeling pipeline.
We construct a corpus of \num{14021} images, classify \num{4831} visualization instances using a hierarchical, domain-informed taxonomy, and analyze patterns of visualization adoption across venues, history subfields, and time.
To interpret these patterns, we conduct interviews with \num{11} historians and use \systemName system as a boundary object to support joint inspection of the corpus.
We identify distinct roles for visualizations in historical research: primary-source, evidence-synthesis, communicative, confirmative, and exploratory.
We further find that while historians pursue diverse goals with figures, persistent epistemological and practical barriers, such as uncertainty, provenance, justification burden, and publication constraints, impede the adoption of visualization.
This work contributes a grounded account of visualization use in historical scholarship and points to opportunities to better support domain-specific needs.


\end{abstract}

\section{Introduction}
\label{sec:introduction}

With the rise of digital humanities (DH) and quantitative history, historians are increasingly using data-driven approaches~\cite{Grandjean2022Data, Drucker2011Humanities}.
As the need to manage and interpret data grows, visualization is increasingly used in historical research for communication, analysis, and knowledge production.
However, a systematic understanding of how historians use visualization is still lacking.

The current understanding of visualization usage in historical scholarship faces two limitations.
First, existing literature and design studies predominantly focus on digital humanities~\cite{Panagiotidou2023Communicating, Jaenicke2017Visual, Bradley2018Visualization}.
While valuable, they represent a technically oriented subset of the field, often overlooking the practices within ``traditional'' history journals where the majority of domain knowledge is produced.
Furthermore, these studies rarely characterize the diversity of visualization styles and functions in research articles.
Second, discussions on visualization usage in historical research remain conceptual~\cite{Drucker2011Humanities,Lamqaddam2018When} or fragmented~\cite{Ewalt2016Image}, focusing on a few exemplary cases.

Without a systematic corpus-driven analysis of visualization usage by historians, the landscape remains unclear.
Consequently, it is difficult to answer the following questions at scale:
Which visualization techniques are effectively adopted in historical research?
How do visualization types distribute over time, publication venues, and history subfields?
Why do historians use visualization, and what are their underlying motivations?
Which visual forms are underused, misused, or largely ignored in historical practice?
What are the potential barriers hindering their adoption?

To address these gaps, we conduct a corpus-driven mixed-methods study of visualization in historical research articles.
First, we construct a three-level corpus from research articles in general history and digital humanities journals, linking articles, extracted images, and visualization instances.
Second, we collaboratively develop a domain-informed visualization taxonomy with historians and scale labels to the full corpus with a semi-automatic pipeline.
Third, recognizing that quantitative data alone cannot explain intent, we adopt a participatory approach involving a boundary object, the \systemName system for exploring the corpus, to facilitate close collaboration with domain experts.
Through interviews with \num{11} historians, we examine how visualizations are interpreted and used in practice, and why certain techniques are ignored or rejected.
This work has the following contributions:

\begin{itemize}
    \item A large-scale corpus of images from history and DH journals, together with a domain-informed taxonomy for visualizations among the images.
    
    \item A mixed-methods workflow that involves corpus labeling and interviews with historians.
    
    \item Empirical findings on the current usage of visualization in historical research articles, historians' motivations and concerns, and patterns of underuse and misuse.
\end{itemize}

The corpus and the \systemName system are available at \url{https://github.com/Sleepydui/HiFigAtlas}.

\section{Related Work}
\label{sec:related-work}

We organize related work into three themes: visualization in historical scholarship (\cref{sec:related-work-hist}), visualization in relation to digital humanities (\cref{sec:related-work-dh}), and visualization taxonomies (\cref{sec:related-work-tax}).

\subsection{Visualization Usage by Historians}
\label{sec:related-work-hist}

Historians have long maintained a divided attitude toward visualization.
While traditional historical scholarship emphasizes careful reading of textual sources, as Martinez described~\cite{Martinez1995Imaging}, digitally oriented approaches embrace visualization as a methodological innovation.
Within historical scholarship, visualizations are increasingly used to support pattern discovery from large-scale materials and to communicate and interpret arguments~\cite{Ewalt2016Image}.
At the same time, critiques warn that visualization can oversimplify complexity~\cite{Correll2019Ethical} and create an illusion of objectivity that discourages critical reflection~\cite{Drucker2011Humanities}.

However, prior discussions remain conceptual and fragmented, grounded in case studies or personal reflection~\cite{Drucker2011Humanities}.
They primarily emphasize bibliographic signals (e.g., keywords, citations, and co-authorship) rather than analyzing visualization content (e.g., chart types and roles).
While some research examines images from perspectives such as art history~\cite{Lamqaddam2018When}, these studies typically focus on historically styled illustrations rather than visualizations authored as analytical or communicative artifacts.
Without corpus-supported quantitative evidence, it is difficult to systematically measure what visualization types historians use, how these patterns vary across venues and topics, and what functions these visualizations serve in practice.
Parallel efforts curate historical visualization collections for dataset-driven analysis~\cite{Zhang2024OldVisOnline,Mei2025ZuantuSet}, but they do not characterize how historians embed and interpret analytic figures in contemporary research articles at scale.
While there have been debates on the pros and cons of visualization in historical scholarship, it remains unclear which types of visualizations trigger these pros and cons, what their functions are, and how misunderstandings and biases can be mitigated.

\subsection{Visualization and Digital Humanities}
\label{sec:related-work-dh}

Interdisciplinary contact between visualization and digital humanities has deepened through venues such as the Workshop on Visualization for Digital Humanities (VIS4DH) at IEEE VIS~\cite{Bradley2018Visualization}.
Digital humanities scholarship has used diverse visualization techniques and has articulated critical perspectives on visualization practice~\cite{DIgnazio2016Feminist,Klein2013Image,Fischer2023Network}.
At the same time, inherent conflicts between visualization and humanities communities arise from epistemological differences~\cite{Panagiotidou2023Communicating}.

A source of tension is the loss of meaning when qualitative humanities data are abstracted into visual encodings~\cite{Lamqaddam2021Introducing}.
This issue leads to the development of models that preserve interpretive context and reframe visualization as an exploratory inquiry where analysts iteratively move between abstractions and source materials~\cite{Hinrichs2016Speculative}.
Furthermore, misalignments in evaluation criteria and power asymmetries among stakeholders in data preparation can hinder effective collaboration~\cite{Janicke2016Valuable, Hinrichs2019Defense}.
These issues manifest as conflicting expectations between technical contributions and domain applicability, as well as the unequal workload borne by humanities experts~\cite{Akbaba2023Troubling}.
These discussions are most developed within DH-oriented venues, while efforts to characterize visualization usage in general history venues remain limited.

\subsection{Visualization Taxonomies}
\label{sec:related-work-tax}

To analyze visualization usage in historical publications, we need taxonomies that categorize visualizations to answer questions such as what types of visualizations are used and in what contexts.
However, existing taxonomies are not directly applicable to visualizations from historical research articles.
Taxonomies for general scholarly publication figures introduced in datasets such as DocFigure~\cite{Jobin2019DocFigure} and ACL-Fig~\cite{Karishma2023ACLFig} target modern scientific figures and are typically coarse-grained.
Visualization taxonomies, such as VIS30K~\cite{Chen2021VIS30K} and VisImages~\cite{Deng2023VisImages}, emphasize novel and composite designs and are not tailored to the historical domain.
Recently, taxonomies specifically oriented toward historical images emerged, such as VisTaxa~\cite{Zhang2025VisTaxa}.
However, VisTaxa's taxonomy is not designed to account for modern charts created by historians.
Moreover, VisTaxa primarily characterizes the visual representation of figures in isolation, whereas we analyze figures within their context (i.e., the research articles) and consider not only their visual representation but also their function in the article and relate them to historians' research practice.

\section{Methodology}
\label{sec:method}

\begin{figure*}
    \centering
    \includegraphics[width=\linewidth]{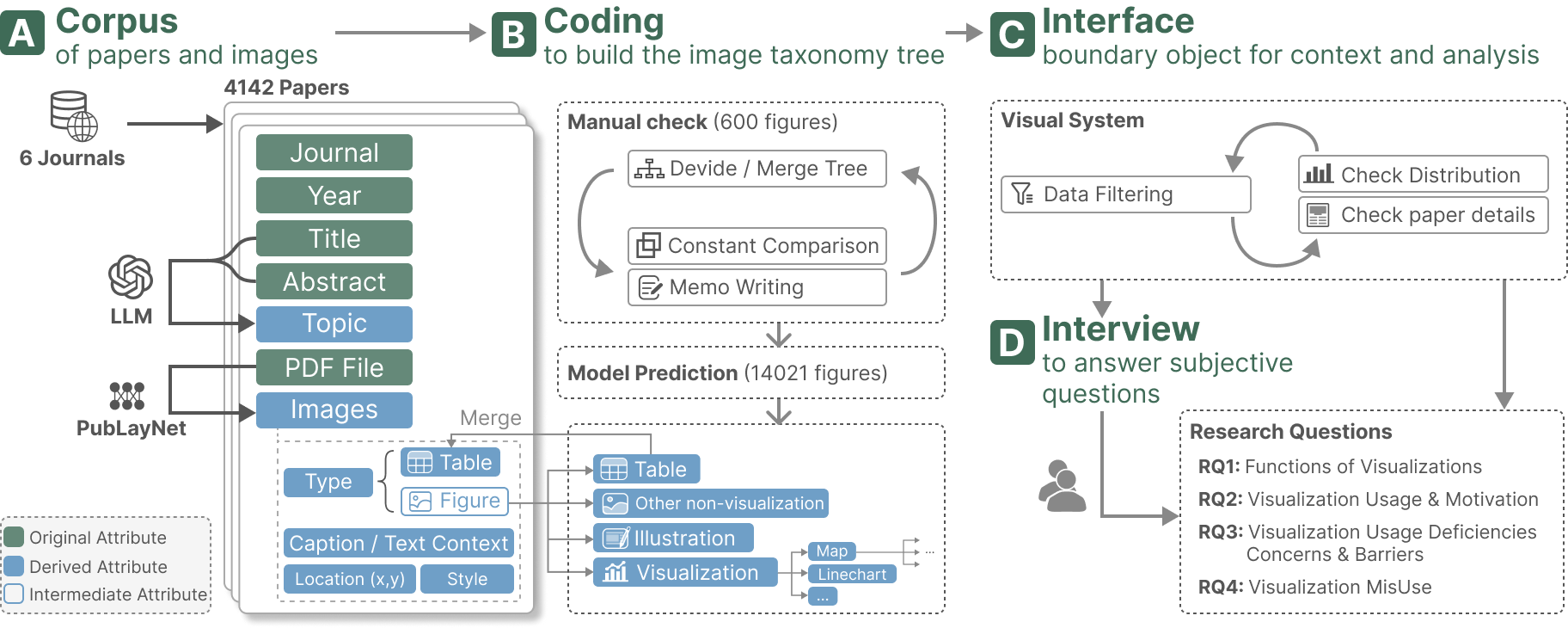}
    \caption{
        Overview of our mixed-methods study workflow.
        (A) We collected \num{4142} articles from six journals, extracted figures from PDFs, and derived contextual metadata (captions and nearby paragraphs).
        (B) We collaborated with historians to code a subset of figures and develop a taxonomy tree within VisTaxa interfaces~\cite{Zhang2025VisTaxa}, and subsequently scaled taxonomy labels to figures in the full corpus.
        (C) The \systemName system was used for collaborating with historians to analyze and validate the classification results.
        The statistical results displayed by the interface, along with
        (D) the results from the interviews with historians, were jointly organized to answer the research questions.
    }
    \label{fig:workflow}
\end{figure*}

We used a corpus-driven mixed-methods approach combining large-scale corpus construction, collaborative taxonomy development, AI-assisted labeling, \systemName as a boundary object, and semi-structured interviews with historians.
This section describes each component of the approach, following the workflow in \cref{fig:workflow}.

\subsection{Corpus}
\label{sec:corpus}

We drew our data from the Social Science Citation Index (SSCI), selecting four top-ranked general history journals classified in SSCI Q1.
We also included two well-regarded digital humanities journals and manually retained only the history-focused articles from them.
We restricted the corpus to research articles, excluding book reviews and other non-research content.
We retrieved DOIs from Web of Science (WoS)~\cite{Clarivate1997Web} and augmented metadata via the Crossref API~\cite{Hendricks2020Crossref}.
Note that our collection is limited by WoS coverage and may not include all articles ever published in the journals.
Following manual verification, \textbf{the final collection comprises \num{4142} research articles}, as shown in \cref{tab:corpus}.

\begin{table}[!htbp]
    \caption{
        Research articles per journal in the final corpus.
    }
    \centering
    \fontsize{7.2pt}{8.6pt}\selectfont
    \begin{tabular}{lrr}
        \toprule
        \textbf{Journal}                      & \textbf{Articles}   & \textbf{Share}    \\
        \midrule
        Past \& Present                       & \num{1403}          & 33.87\%           \\
        \midrule
        The American Historical Review        & \num{1328}          & 32.06\%           \\
        \midrule
        Historical Methods                    & \num{598}           & 14.44\%           \\
        \midrule
        History and Technology                & \num{322}           & 7.77\%            \\
        \midrule
        Digital Scholarship in the Humanities & \num{252}           & 6.08\%            \\
        \midrule
        Digital Humanities Quarterly          & \num{239}           & 5.77\%            \\
        \midrule
        \textbf{Total}                        & \textbf{\num{4142}} & \textbf{100.00\%} \\
        \bottomrule
    \end{tabular}
    \label{tab:corpus}
\end{table}

For each article, we used PubLayNet~\cite{Zhong2019PubLayNet} to detect and extract page-level graphical elements and assign initial labels (figure or table).
We then used Tesseract~\cite{Smith2007Overview} for OCR and applied spatial heuristic rules (including proximity, column alignment, and caption keyword matching) to extract nearby paragraphs as context for each figure and table.
This process resulted in \textbf{a corpus of \num{10176} figures and \num{3845} tables}.
Note that during processing, we do not divide figures.
A figure may contain multiple subfigures.

\subsection{Coding Process and Taxonomy Construction}
\label{sec:coding}

Building on the corpus described in \cref{sec:corpus}, we categorized the \num{10176} extracted figures to develop a taxonomy that is meaningful to both historians and visualization researchers.
We adapted the coding protocol of VisTaxa~\cite{Zhang2025VisTaxa}: a taxonomy is iteratively developed through manual annotation of a small subset, then scaled to the full corpus via model prediction and expert validation.
During manual coding, we used a human-AI collaborative pipeline that integrates cross-disciplinary co-coding, constructivist grounded theory~\cite{Charmaz2024Constructing}, and machine assistance.

Two of the authors, a historian (C1) and a visualization researcher (C2), worked closely as coders using VisTaxa's coding interfaces.
They iteratively refined categories through constant comparison, memo writing, and discussion of edge cases.
The coders used \systemName (\cref{sec:system}) to inspect figures together with their surrounding page, caption, and nearby paragraphs.

Six disjoint batches of images were randomly sampled from the dataset, each with 100 images.
We refer to these batches as B1--B6.
Each batch followed a four-step protocol:

\begin{enumerate}[leftmargin=5mm]
    \item[S1] \textbf{Draft taxonomies and initial labels (individual).}
          Each coder independently drafts a hierarchical taxonomy and assigns multi-label annotations to the same set of images.
          The labels include both visualization \emph{type} and \emph{role} (discussed in \cref{sec:functions}).

    \item[S2] \textbf{Resolve structural conflicts (group).}
          The coders compare their taxonomies and consolidate a shared taxonomy.

    \item[S3] \textbf{Refine labels (individual).}
          Each coder relabels images against the consolidated taxonomy.

    \item[S4] \textbf{Resolve label conflicts (group).}
          The coders review images with conflicting labels, aligning definitions and interpretations.
\end{enumerate}

The coders are allowed to keep divergent opinions on the taxonomy and labels.
The steps are repeated in batches until convergence, where no new categories emerge.

In our practice, no new taxa emerged in B5 and B6.
The final taxonomy consists of \nTaxaFinalAllVis taxa (excluding the \taxon{root}, \taxon{non-visualization}, and its subcategories).
Representative images and definitions for each taxon are provided in \iflabelexists{appendix:taxon-definitions}{\cref{appendix:taxon-definitions}}{the ``Taxon Definitions'' section in the supplemental material}.
We used Krippendorff's alpha with MASI distance to evaluate inter-coder reliability for the labels.
Results showed high consistency in later batches: B4 achieved 388/400 exact matches ($\alpha=0.79$), and B6 achieved 559/600 exact matches ($\alpha=0.74$).
The discussions to resolve conflicts (S4) do not force full consensus.
The final unresolved edge cases (41/600 of the coded dataset) were marked as ambiguous and archived.

A key decision in the coding process is the boundary between visualization and illustration, as the two lie on a continuum.
The coders converged on working definitions: a \emph{visualization} is ``a visual representation that commonly uses visual design to represent data,'' whereas an \emph{illustration} is ``a non-visualization that commonly uses drawings, sketches, or paintings to provide visual explanations of concepts, plans, processes, or scenes'' (see \iflabelexists{appendix:taxon-definitions}{\cref{appendix:taxon-definitions}}{the ``Taxon Definitions'' section in the supplemental material}).
Maps, which encode geographic information through symbols and color, are classified as visualizations under this criterion.
Even sketch maps or painterly maps qualify as long as they encode locational data.
Meanwhile, some instances remain genuinely ambiguous.
For example, a comic of an imagined view of the moon may be read by a coder with visualization background as a scene without encoded data, and thus as an illustration, whereas a coder without visualization background may argue that it encodes imagined lunar terrain, making it a visualization.
We consider both readings defensible and retained such disagreements as unresolved ambiguous cases.

To scale the taxonomy labels from the 600 coded samples to the full corpus, we formulated labeling as a hierarchical multi-label classification problem and compared two strategies provided by VisTaxa~\cite{Zhang2025VisTaxa}.
\emph{Similarity-based matching} extracts CLIP ViT-B/32~\cite{Radford2021Learning} embeddings for each image and transfers the labels of the nearest labeled neighbor by cosine similarity.
\emph{Zero-shot classification} prompts Qwen3-VL~\cite{Bai2025Qwen3VL} with the names and hierarchical structure of our taxonomy.
The model predicts one or more leaf categories per image, which are then lifted to ancestor nodes to form a full hierarchical label set.

We evaluated both methods on the 600 expert-labeled images using exact-match ratio and Jaccard similarity (excluding the root), each reported at both the full label depth (marked as $Full$) and at the top level only (marked as $D{=}1$).
\Cref{tab:evaluation} summarizes the results.
Similarity-based matching achieves higher full-depth agreement, whereas zero-shot classification performs substantially better at the top level.
Both methods produced similar corpus-level estimates of the number of visualizations (\num{4960} versus \num{4831}).

\begin{table}[!htbp]
    \caption{
        Comparison of similarity-based matching and zero-shot classification for label prediction on \num{600} expert-labeled images and the number of estimated visualizations (\#VIS) in the corpus.
    }
    \centering
    \fontsize{7.2pt}{8.6pt}\selectfont
    \begin{tabular}{p{0.20\linewidth}ccccc}
        \toprule
        \multirow{2}[1]{*}{\textbf{Method}}  & \multicolumn{2}{c}{\textbf{Exact match}} & \multicolumn{2}{c}{\textbf{Jaccard}} & \multirow{2}[1]{*}{\textbf{\#VIS}}                               \\
        \cmidrule(lr){2-3}\cmidrule(lr){4-5} & \textbf{Full}                            & $\mathbf{D=1}$                       & \textbf{Full}                      & $\mathbf{D=1}$ &            \\
        \midrule
        similarity-based                     & 53.33\%                                  & 76.83\%                              & 0.635                              & 0.804          & \num{4960} \\
        zero-shot                            & 45.00\%                                  & 86.67\%                              & 0.619                              & 0.897          & \num{4831} \\
        \bottomrule
    \end{tabular}
    \label{tab:evaluation}
\end{table}

In addition to the algorithmic evaluation, we further validated the quality of the predicted labels by manually evaluating a random sample of 100 images and their predicted labels, judging whether the predicted top-level and leaf-level labels were correct, partially correct, or incorrect.
The assessment was conducted by the two coders independently and then aligned through discussion.
Of the 100 images, 89 had correct top-level predictions and 55 had fully correct leaf-level predictions, confirming that top-level distinctions are substantially more stable than fine-grained leaf assignments and that model errors concentrate among closely related leaf categories.
Manual validation revealed complementary error patterns: 

\begin{itemize}
    \item Similarity-based matching sometimes transferred overly specific or semantically incompatible multi-label sets from a visually similar but conceptually unrelated image.
    \item Zero-shot classification tended to confuse only closely related sibling categories, producing errors that were easier to detect and more consistent with expectations of historians involved in the interview at the top level.
\end{itemize}

Since our downstream analyses rely primarily on top-level distinctions, we adopted zero-shot classification for generating labels used in the remainder of this study.

\subsection{\systemName System as a Boundary Object}
\label{sec:system}

\begin{figure*}
    \centering
    \includegraphics[width=1\linewidth]{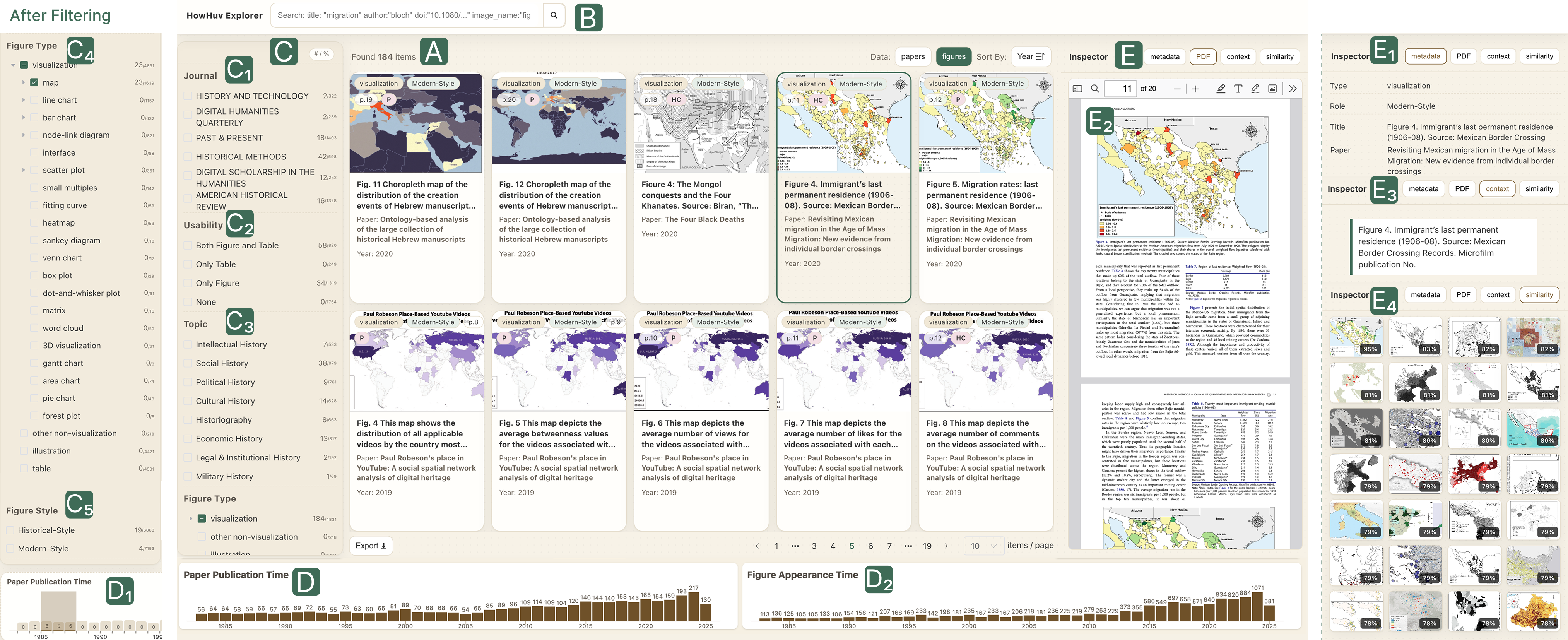}
    \caption{
        \systemName system, which serves as a boundary object.
        (A) displays the list of figures or papers;
        (B)--(D) provide multi-dimensional filtering capabilities, where (B) is a search bar supporting full-text retrieval across titles, abstracts, and DOI fields;
        (C) enables compound filtering based on multiple criteria, including (C1) journal sources, (C2) whether an article contains visualizations or tables, (C3) article topics, and (C4) chart types;
        (D) offers time-based brushing functions that show the distributions of article counts (D1) and chart counts (D2) across different periods.
        (E) presents detailed information about the selected chart or article, including four sections: (E1) metadata, (E2) PDF, (E3) the textual context in which the chart appears, and (E4) similarity.
    }
    \label{fig:interface}
\end{figure*}

Our interview study (\cref{sec:interview}) required historians and visualization researchers to reason about the same figures, labels, and patterns while bringing different disciplinary vocabularies and expectations.
We therefore designed \systemName as a web-based boundary object over our corpus (\cref{fig:interface}).
\emph{Boundary object} refers to an interactive artifact that supports cross-disciplinary collaboration by maintaining a shared reference point while remaining open to reinterpretation and negotiation~\cite{Star2010This,Lee2007Boundary}.

\systemName exposes the full image corpus, metadata, taxonomy, and label provenance in coordinated views.
For historians, it preserves familiar practices of reading research articles and scanning figures in their original context.
For visualization researchers, it provides structured filters, distributions, and similarity views aligned with the taxonomy.
The goal is not to enforce an authoritative classification but to provide a common environment in which disagreements can be inspected, negotiated, and documented.
For example, when a historian and a visualization researcher disagreed on the label for the same figure, both labels and their provenance were shown in the same inspector panel.
The two experts could then compare the figure's surrounding text, publication context, and visually similar cases before revising or retaining their judgments.

Users begin with the list of papers or figures (\cref{fig:interface}(A)), and progressively narrow the corpus through cross-filtering (\cref{fig:interface}(B)--(D)).
Filters across different components are combined conjunctively, while multiple selections within the same component are aggregated disjunctively, enabling flexible yet precise exploration.
Selecting an item opens the detail inspector (\cref{fig:interface}(E)), where users can switch among metadata, the original PDF page, contextual excerpts, and similarity view (\cref{fig:interface}(E1)--(E4)).
The similarity view connects the selected figure to visually related figures, supporting associative exploration beyond direct filtering.

\systemName played three methodological roles:

\begin{enumerate}[leftmargin=*]
    \item \textbf{Grounded taxonomy alignment.}
          During the coding process (\cref{sec:coding}), ambiguous or contested cases were loaded into the inspector together with the PDF and surrounding text.
          Coders discussed them until definitions and label boundaries converged.

    \item \textbf{Pattern and blind-spot discovery.}
          Using filters and temporal views, we identified clusters of frequently used forms, underused opportunities, and potential misuses, which then informed our quantitative analyses and interview prompts.

    \item \textbf{Evidence capture.}
          During coding (\cref{sec:coding}) and interviews (\cref{sec:interview}), decisions, edge cases, and alternative readings discussed around specific figures were recorded as memos linked to those items.
\end{enumerate}

\subsection{Semi-Structured Interviews with Historians}
\label{sec:interview}

To understand the motivations behind historians' visualization practices and to cross-validate and enrich findings in our corpus, we conducted semi-structured interviews with \num{11} historians (\cref{tab:interviewees}).
Interviewees spanned academic ranks, history subfields, and self-reported experience with visualization and digital humanities.

\textbf{Procedure.}
Each interview, conducted online or in person, lasts 60 -- 120 minutes and covers four topics (detailed in \iflabelexists{appendix:interview-procedure}{\cref{appendix:interview-procedure}}{the ``Interview Procedure'' section in the supplemental material}):

\begin{enumerate}[leftmargin=*]
    \item Research background and prior experience;
    \item Use, functions, and perceived value of visualization;
    \item Corpus and taxonomy-based exploration with \systemName;
    \item Nonuse, along with associated risks, and concerns.
\end{enumerate}

\textbf{Analysis process.}
We analyzed interview data using codebook thematic analysis~\cite{Braun2021Thematic}.
Two authors independently read the transcripts for familiarization and noted initial observations.
Guided by the four research questions listed in \cref{fig:workflow}(D), we then collaboratively developed an initial codebook, applied and refined codes, and grouped related codes into higher-level themes.
Throughout this process, we cross-checked each other's coding and theme definitions, documented disagreements, and iteratively revised the codebook to reach a shared understanding of theme boundaries.
In reporting, we refer to interviewees as H1--H11.

\newlength\idw
\newlength\sw
\newlength\lw
\setlength{\idw}{0.02\linewidth}
\setlength{\sw}{0.15\linewidth}
\setlength{\lw}{0.22\linewidth}
\begin{table}[!htbp]
    \caption{
        Interviewee demographics, covering academic position, specialization, years of research experience, and self-reported experience with visualization.
    }
    \centering
    \fontsize{7.2pt}{8.6pt}\selectfont
    \begin{tabular}{@{}p{\idw}p{0.18\linewidth}p{\lw}p{\sw}p{\lw}@{}}
        \toprule
        \textbf{ID} & \textbf{Position}    & \textbf{Specialization}          & \textbf{Year of\ Experience} & \textbf{VIS\ Experience} \\
        \midrule
        H1          & Associate Professor  & Ancient History                  & $>$\,20              & Frequent use               \\
        \midrule
        H2          & PhD Candidate        & Philology                        & 6--10                & Frequent use               \\
        \midrule
        H3          & Professor            & World History                    & $>$\,20              & Occasional use             \\
        \midrule
        H4          & Professor            & History of Science \& Technology & $>$\,20              & Occasional use         \\
        \midrule
        H5          & PhD Candidate        & Intellectual History             & 6--10                & No prior experience        \\
        \midrule
        H6          & Lecturer             & Early Modern History             & 16--20               & Occasional use         \\
        \midrule
        H7          & Associate Professor  & Social \& Intellectual History   & 16--20               & Frequent use               \\
        \midrule
        H8          & PhD Candidate        & Philology                        & 6--10                & No prior experience        \\
        \midrule
        H9          & PhD Candidate        & History of Science \& Technology & 6--10                & Occasional use             \\
        \midrule
        H10         & PhD Student          & Historical Geography             & 6--10                & Frequent use               \\
        \midrule
        H11         & Postgraduate Student & World History                    & 5                    & Frequent use               \\
        \bottomrule
    \end{tabular}
    \label{tab:interviewees}
\end{table}

\section{Taxonomy and Distribution}
\label{sec:taxonomy}

This section summarizes corpus-level patterns after applying the taxonomy and labeling pipeline described in \cref{sec:coding}.
Our analysis draws on \num{4142} research articles from six history and digital humanities journals (\cref{sec:corpus}).
From the \num{14021} extracted figures and tables we identified \num{4831} visualization instances, \num{4501} \taxon{table}s, \num{4471} \taxon{illustration}s, and \num{218} other \taxon{non-visualization}s.
Category definitions appear in \iflabelexists{appendix:taxon-definitions}{\cref{appendix:taxon-definitions}}{the ``Taxon Definitions'' section in the supplemental material}.
The visualization subset (i.e., excluding excluding the \taxon{root}, \taxon{non-visualization}, and its subcategories) is described by \nTaxaFinalAllVis taxa in our hierarchy, among which \nTaxaFinalFirstLevelVis are at the first level below the root and \nTaxaFinalLeafVis are leaves (\iflabelexists{appendix:visualization-taxonomy}{\cref{appendix:visualization-taxonomy}}{see the ``Visualization Taxonomy'' section in the supplemental material}).
Linking articles, images, and taxa and browsing them in \systemName lets us characterize what historians visualize, how often, and how usage varies by venue, subfield, figure style, and chart type.

\textbf{Temporal adoption.}
The number of visualization instances grows with the size of the literature.
Visualization appears in a steadily larger share of articles: \percentageWithRawData{19.64\%}{11/56} in 1983, \percentageWithRawData{30.88\%}{21/68} in 2003, and \percentageWithRawData{41.94\%}{91/217} in 2024.
This pattern suggests a gradual normalization of figures in journals rather than a single abrupt ``visual turn''.

\textbf{Historical-style versus modern-style images.}
We distinguish \emph{historical-style} images, typically pre-digital artifacts that may function as primary evidence, from \emph{modern-style} images produced by authors as secondary representations (\cref{sec:functions} elaborates on this distinction).
Using Qwen3-VL~\cite{Bai2025Qwen3VL} for zero-shot classification with manual spot checks, we observed \num{6868} historical-style and \num{7153} modern-style images among the \num{14021} figures and tables.
The model prompt was: ``Please determine whether this image is an old visualization (primary source) or a historian-created visualization (secondary source).''

\textbf{Venues.}
Adoption is uneven across journals.
\percentageWithRawData{84.52\%}{213/252} of the articles in \journalName{Digital Scholarship in the Humanities} include at least one visualization, compared with 54.81\% in \journalName{Digital Humanities Quarterly}.
Among general history venues, only \journalName{Historical Methods}, with its methodological focus, approaches a similar profile (\percentageWithRawData{59.87\%}{358/598} of articles with visualizations).
The other three journals' visualization adoption rates remain below 30\%.

\textbf{History subfields.}
We grouped articles into eight subfields using the index of topics from \journalName{The American Historical Review}~\cite{TheAmericanHistoricalReview2019Index} (see \iflabelexists{appendix:subfields}{\cref{appendix:subfields}}{the ``Article Subfields and Topics'' section in the supplemental material}).
Economic history is the most visualization-intensive subfield: \percentageWithRawData{41.96\%}{133/317} of the articles contain at least one visualization, and the dominant chart type is the \taxon{line chart} (\percentageWithRawData{37.42\%}{183/489}).
In historiography, \taxon{node-link diagram} is most prevalent (\percentageWithRawData{30.94\%}{306/989}), which may be related to its focus on the relationships between concepts and words.
In the remaining subfields, the most frequent visualization type is the \taxon{map}.

\textbf{Chart-type concentration.}
A small set of chart types dominates the visualization usage.
\taxon{Map} accounts for \percentageWithRawData{33.93\%}{1639/4831} of all visualization instances and spans \num{12} distinct leaf categories, the richest branch of the taxonomy tree.
Beyond maps, historians rely heavily on tables and statistical graphics.
\taxon{Line chart} (\percentageWithRawData{23.95\%}{1157/4831}) and \taxon{bar chart} (\percentageWithRawData{13.08\%}{632/4831}) are frequently used to summarize counts, shares, or trends.
These charts often facilitate comparative analysis.
Among \num{1157} \taxon{line chart}s and \num{632} \taxon{bar chart}s, grouped configurations appear \num{114} and \num{148} times, respectively.
We observe \num{142} \taxon{small multiples} where visualizations are used for side-by-side comparison.
The corpus also contains \num{88} \taxon{interface} images (screenshots of analytical tools, database front ends, and dashboards) which emerged relatively late.
The earliest instance the corpus records is an ArcMap~\cite{Esri1999ArcMap} interface in \journalName{Historical Methods} in 2007~\cite{St-Hilaire2007Geocoding}.
Following Rosenberg and Grafton~\cite{Rosenberg2010Cartographies}, we use \emph{temporal-structure visualizations} to refer to encodings whose primary metaphor is time as structure, such as timelines, Gantt charts, and storylines.
Under this definition, statistical charts with time on an axis are not temporal-structure visualizations.
Despite the importance of time in historical research, temporal-structure visualizations are scarce in the corpus.
Storylines are almost absent, and \taxon{Gantt chart}s account for only \percentageWithRawData{0.06\%}{3/4831} of visualization instances.

In summary, visualization usage is widespread but uneven: \taxon{map}, \taxon{node-link diagram}, and \taxon{line chart} are entrenched and align with data that historians view as routine and defensible.
These corpus patterns align with historians' self-reported practices from the interviews, with the high prevalence of \taxon{map} and \taxon{node-link diagram} matching the actual usage reported by historians (\cref{sec:functions,sec:motivations}).
Building on this empirical foundation, the subsequent sections explore the functional roles of visualization, historians' motivations, and the factors impeding the adoption of specific visual forms.

\section{Functions of Visualizations in Historical Research}
\label{sec:functions}

General visualization theory defines abstract tasks such as filtering, sorting, and deriving~\cite{Munzner2014Visualization}.
This atomic perspective does not capture the workflows of historical research, spanning evidence collection, source criticism, and narrative construction, nor the distinctive interactions between researchers and primary or secondary sources~\cite{Collingwood1994Idea}.
Existing classifications of visualization usage in historical research are either too coarse, such as the binary distinction between ``infographics'' and ``data visualization''~\cite{Grandjean2022Data}, or too narrow, such as focusing on uncertainty visualization~\cite{Panagiotidou2023Communicating}.
Neither comprehensively describes the broad functions of visualization instances observed across history research articles.

To address this gap, we developed five functional roles through iterative synthesis across two complementary evidence sources.
During taxonomy construction (\cref{sec:coding}), we observed recurring image patterns, e.g., primary-source visualizations with distinctive historical styles and visual analytics system screenshots that appear to support exploratory analysis.
Because historians rarely articulate intended functions explicitly in writing, we treated these observations as provisional hypotheses.
We then probed these hypotheses in the interviews (\cref{sec:interview}), letting historians elaborate on how visualizations are actually interpreted and used in practice.
The five roles are not mutually exclusive and are presented below.

\textbf{Role 1: Primary-Source Visualization} denotes visual forms directly inherited from original sources rather than generated by historians, serving primarily as objects of study or historical evidence.
H2 noted that certain history subfields treat ``images as historical evidence''~\cite{Burke2001Eyewitnessing}, and H9 classified such images as primary sources.
These historical images are visually and materially distinct from modern-style visualizations and encompass maps, star charts, or calendars found in archives, manuscripts, and artifacts.
H6 emphasized the necessity of analyzing elements within war maps through documentary interpretation rather than equating them with modern charts.
H10 noted that historians also extract data from these sources.
An example in our corpus is \cref{fig:case}(A), which presents Buchholz's 1956 epidemic map~\cite{Mengel2011Plague}, cited by the history paper author to illustrate how maps influence varying understandings of the Black Death's transmission.
It differs from modern-style visualizations in font, printing clarity, and paper texture.

\begin{figure}
    \centering
    \includegraphics[width=1\linewidth]{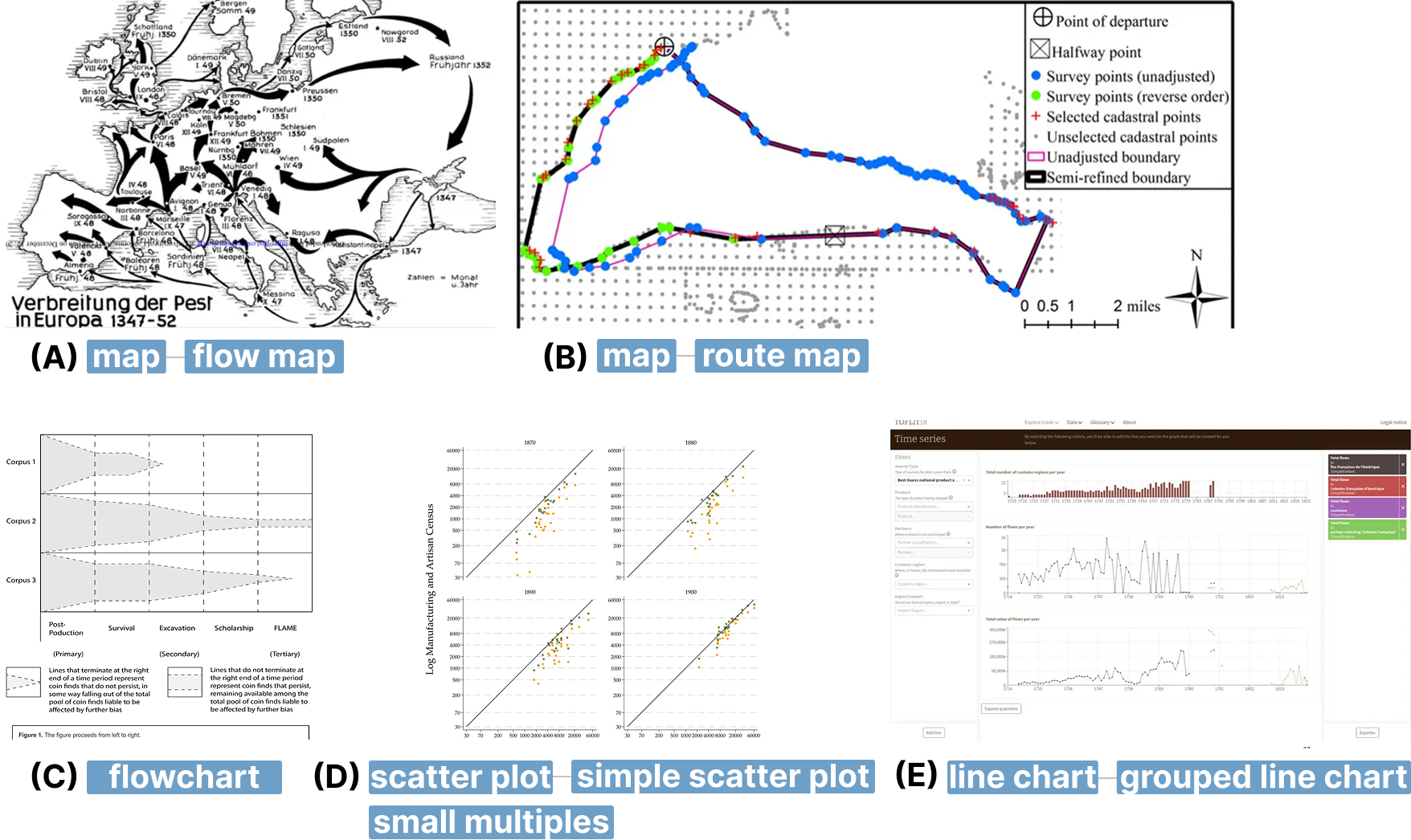}
    \caption{
        Example visualization types in our corpus.
        (A) \taxon{Flow map}~\cite{Mengel2011Plague}.
        (B) \taxon{Route map}~\cite{Storey2019Cartographically}.
        (C) \taxon{Flowchart}~\cite{Mordechai2025Systematic}.
        (D) \taxon{Simple scatter plot} and \taxon{small multiples}~\cite{Almas2026Historical}.
        (E) \taxon{Grouped line chart}~\cite{Charles2022Exploring}.
    }
    \label{fig:case}
\end{figure}

\textbf{Role 2: Evidence-Synthesis Visualization} refers to visualizations generated by historians, either manually or with tools such as Gephi~\cite{Bastian2009Gephi} and QGIS~\cite{Dawson2026Qgis}, to aggregate heterogeneous and scattered primary sources.
Because historians prioritize organizing original materials, these visualizations primarily serve as workbenches that facilitate the arrangement of clues and the critical analysis of evidence.
For instance, H8 integrated interpersonal relationships from dispersed archives using genealogies and tables.
Such images typically function as intermediate artifacts during the research process and are rarely published in research articles.
The semi-refined reconstruction of the Las Truchas land boundaries in \cref{fig:case}(B) is an exception~\cite{Storey2019Cartographically}, included specifically to illustrate data processing methodologies.
Interviews indicate that although such visualizations are rare in publications, they are critical in early research stages.
Almost all interviewees (H1--H11) employ such methods when organizing materials: tables often serve as the initial step in constructing visualizations (H2, H3, H4), while H1 and H6 construct maps through manual compilation.
Even H5, who claimed no visualization experience, used color-coded multidimensional tables to organize correspondence between historical individuals.
Furthermore, scholars use complementary visualizations alongside original historical data, such as the open-source maps used by H2, to enhance comprehension, a practice H2 described as ``understanding the person in their context.''

\textbf{Role 3: Communicative Visualization} is the most prevalent category, used by historians to substantiate arguments or disseminate findings.
These visualizations aim to present established conclusions intuitively rather than support open-ended exploration.
Common forms include bar charts for summarizing frequencies, line charts for displaying trends, and heatmaps for representing regional intensities.
For instance, the flowchart in \cref{fig:case}(C) illustrates the quantitative attrition of excavated coin data across successive stages~\cite{Mordechai2025Systematic}.
These images are typically used after the analysis is completed.
H3 noted that bar charts clearly reveal disparities, while H5 pointed out that visualization can ``give shape'' to abstract arguments.
Beyond research articles, historians use communicative visualizations in presentations and on social media to lower the barrier to understanding (H2, H4, H9).
Furthermore, they enhance methodological transparency by demonstrating research workflows, as evidenced by the \num{404} flowcharts in the corpus.

\textbf{Role 4: Confirmative Visualization} aims to model and explore the uncertainty inherent in historical knowledge.
By highlighting historical contingency and evidentiary gaps, these visualizations are employed to confirm or refute hypotheses and to present multiple interpretive paths.
Forms include fitting curves, probability estimates, and counterfactual comparisons.
A classic example is Robert Fogel's work in economic history~\cite{Fogel1995Time}, which used scatter plots and fitted lines to model the distribution of male slave prices by age in the Old South.
Our corpus contains similar instances: \cref{fig:case}(D) uses small multiples to compare the consistency between manufacturing census and artisan census records~\cite{Almas2026Historical}, enabling readers to assess the reliability of one source against another.

\textbf{Role 5: Exploratory Visualization} functions as a knowledge production tool designed to facilitate pattern discovery through interactive means.
These visualizations frequently integrate dashboards or visual analytics systems that support multi-scale transitions between distant reading and close reading.
Interviews reveal that historians regard them as essential tools for thought (H7).
H4 and H2 expressed expectations for using visual analytics to identify anomalies or data outliers, while H8 used network visualization to detect unexpected communities.
Although H8 had not directly applied such methods, after examining the TOFLIT18 project~\cite{Charles2022Exploring} (\cref{fig:case}(E)) within \systemName, they noted that interactive systems combining relational exploration with data querying would significantly enhance verification efficiency.

\section{Motivations and Benefits Reported by Historians}
\label{sec:motivations}

This section summarizes historians' motivations for using visualization: managing scale, stabilizing relational structures, communicating claims, and demonstrating methodological accountability.

\textbf{Addressing data scale.}
Faced with voluminous archives and secondary literature, interviewees emphasized the difficulty of cognitively processing large-scale evidence without distortion.
Visualization supports compression with accountability: counting, summarizing, and comparing at a scale that would otherwise be error-prone or infeasible, while keeping the aggregation logic inspectable.
For example, H7, working with the Chinese Biographical Database (CBDB)~\cite{Tsui2020Harvesting}, relies on summary statistics and network graphs to avoid over-attending to a few prominent individuals or canonical narratives.
H11 similarly framed visualization as an external aid to memory, helping maintain a stable overview when reasoning across many documents.
Compared with close reading alone, the practices of quantification and visualization make analytical steps more explicit and reproducible, alleviating selection bias and human memory limitations.

\textbf{Externalizing relational structures.}
Historians often reason about entangled relationships among people, places, texts, events, and concepts that are difficult to trace through linear prose.
Node-link diagrams and trees help stabilize relational structures into a view that can be revisited, revised, and interrogated.
H8, for instance, used family trees to aggregate scattered kinship and generational information.
The tree visualization made lineage and marriage alliances visible and easier to compare across cases.
Similarly, H2 used flowcharts to elucidate pathways of phonological evolution.
In these accounts, networks and trees were not merely presentational devices but working representations that keep multiple pathways simultaneously available for historical reasoning.

\textbf{Improving expressiveness and persuasiveness.}
Many visualization instances in publications serve communicative purposes (\cref{sec:functions}).
Historians described using simple statistical charts to explain claims efficiently and to reduce interpretive divergence between author and reader.
For example, H3 noted that bar charts and line charts can condense textual descriptions into intuitive graphical representations, making trends and extremes immediately visible.
At the same time, interviewees were acutely aware that this explanatory power entails rhetorical weight: a chart can appear more definitive than the underlying evidence warrants, requiring careful design choices.

\textbf{Reinforcing methodological legitimacy.}
Beyond communicating outcomes, historians also visualize data and evidence processing workflows to enhance transparency, legitimacy, and provenance.
H5 and H7 regarded traceability as non-negotiable: elements in a visualization must remain traceable to original sources, or the visualization loses historiographical value.

Taken together, these motivations clarify why historians prefer visualization forms that are legible, revisitable, and accountable, and why the same properties become points of friction when the data are ambiguous, the visual rhetoric is too strong, or publication constraints limit expressiveness.
We turn next to these concerns and practical barriers in \cref{sec:concerns}.

\section{Concerns and Barriers Reported by Historians}
\label{sec:concerns}

Although historians recognize clear benefits of visualization (\cref{sec:motivations}), the interviews reveal a set of interrelated concerns that impede the adoption of visualization in research articles.
The concerns are rooted in disciplinary epistemology, data characteristics, and practical constraints.

\textbf{Why temporal-structure visualizations are rare.}
\Cref{sec:taxonomy} shows that temporal-structure visualizations (e.g., timelines and storylines) have a notably low adoption rate despite temporality being central to historical research.
Interviewees offered three explanations.
First, \emph{text is often considered sufficient}.
H2 argued that determinate temporal information can be adequately conveyed in text, making charts unnecessary.
H6 commented that in history publications, the relevant temporal information is usually disciplinary common knowledge and thus does not require visualization.
Second, \emph{scale and scope of historical inquiry influence adoption}.
H5 and H8 noted that extended timelines spanning decades or centuries are more typical of public history, whereas research articles tend to focus on narrow temporal windows where timelines provide limited analytical value.
Third, and most fundamentally, \emph{historical temporality resists linear representation}, making conventional temporal-structure visualizations inappropriate.
Recurring cycles (e.g., economic or political), overlapping durations, and intellectual undercurrents do not map cleanly onto a single axis.
They involve temporal ambiguity, multi-scale dynamics, and partially ordered relationships that resist precise placement along a unified timeline.
As H7 emphasized, imposing a linear timeline can introduce a false sense of coherence, making events appear more sequentially ordered than the underlying evidence supports.
This risk of distortion leads historians to favor alternative representations, such as narrative accounts, tables, or small multiples, which better accommodate overlap, uncertainty, and heterogeneous temporal scales.

\textbf{Concerns about data: processing, reduction, and augmentation.}
Visualization in historical research typically sits at the end of a data pipeline, and concerns accumulate at each stage, leading historians to avoid complex visualizations.
In data processing, H11 described a legitimacy dispute around computational methods: using AI techniques such as natural language processing to analyze historical archives necessitates extensive methodological justification, and because visualization follows these steps, it inherits a substantial burden of proof.
In data reduction, H4 highlighted the loss of a ``sense of meaning'' when algorithmic clustering and dimensionality reduction compress rich sources into visual summaries.
H2 maintained that visualizations reliant on complex algorithmic choices should not be published unless the algorithms' applicability can be clearly elucidated.
In data augmentation, H8 and H6 criticized the quantification of social relationships in network visualizations (e.g., defining tie strength by the frequency of dining together), arguing that this practice introduces controversial social models disguised as objective facts.
Regarding the use of environmental data, H10 noted that while continuous phenomena such as temperature are acceptable if the interpolation methodology is transparent, politically defined facts such as historical borders ``cannot be filled'' without risking the fabrication of evidence.

\textbf{Concerns about expressiveness.}
Historians share a widely held concern that standard visual forms lack the expressiveness to support historical reasoning.
Constrained by page and screen dimensions, scholars are often forced to choose between overview and detail.
H4 noted that thematic maps related to the history of mathematics ``omitted excessive detail,'' while a node-link diagram with over 300 mathematicians led him to question the representational limits of the form.
Encoding channels of common visualization types are also perceived as restrictive.
H2 expressed a dislike for Sankey diagrams, arguing that while their fixed hierarchy is effective for tree structures, it cannot accommodate relationships among multiple entities.
Furthermore, static media struggle to represent temporal evolution.
H10 noted that current publishing platforms lack support for dynamic displays, and found it difficult to see how static visualizations could simultaneously depict the movement of individuals across locations in complex temporal sequences.
Beyond these specific limitations, after exploring the corpus in \systemName, H7 expressed disappointment that visualization has not yet genuinely evolved into a driver of ``knowledge production''.
H10 echoed this skepticism, asking, ``Use visual analytics systems, and then what?''
While she anticipated that exploratory systems would stimulate historiographical theoretical innovation, she noted that existing tools fail to effectively support this objective.
Conversely, H2 maintained an optimistic stance, regarding the widespread use of node-link diagrams in digital humanities for analyzing interpersonal relationships as a significant innovative practice.

\textbf{Concerns about subjectivity.}
H9 commented that history is not an exact science, as narrative and model construction inherently entail incompleteness and imagination.
Visualization encounters resistance not due to the inherent subjectivity of data, but because \emph{it tends to conceal subjective choices} more effectively than footnotes or text.
H2 and H7 emphasized that images carry strong rhetorical authority, leading readers to perceive visual encodings, such as continuous scales or clear boundaries, as indisputable facts.
H2 was particularly cautious about mapping uncertain data onto seemingly precise graphics, warning that readers may misinterpret them as settled reality while overlooking underlying ambiguities.
A notable tension exists regarding the attribution of subjectivity.
H11 viewed large-scale computation as a mechanism to mitigate human bias, yielding ``rationally processed sources.''
Conversely, H5 and H8 argued that algorithmic subjectivity embedded in training data and model architectures is more opaque and harder to interrogate than the explicit interpretive stances of historians.
H7 offered an insight regarding this issue: simplification is inevitable in any medium of communication.
The critical challenge lies \emph{not in avoiding simplification but in maintaining reflexive awareness of where and how it occurs}.
Visualization will gain acceptance in the historical community only when it supports, rather than obscures, this reflexivity.

\textbf{Practical barriers: usability, cost, literacy, and publication constraints.}
Alongside epistemological concerns, practical barriers constrain visualization adoption.
Existing tools often present prohibitive technical barriers.
H9 criticized professional visualization tools for presenting excessive parallel functions that obscure entry points.
While powerful, tools such as QGIS and Gephi lack usability tailored to historical research (H7), and their steep learning curves drive scholars toward Excel or hand-drawn sketches that integrate seamlessly into established habits of reading, note-taking, and writing.
A deficit in visualization literacy constitutes another bottleneck.
H7 and H11 acknowledged limited familiarity with visualization techniques, frequently discovering suitable yet previously unencountered visualization forms while browsing the corpus in \systemName.
Concerns regarding visualization literacy of audiences and peers further reinforce conservatism, prompting H3 to prioritize simple bar charts to align with readers' cognitive expectations and minimize ambiguity.
H10 noted that constructing publicly available interactive systems is often perceived as an extraneous burden beyond core research.
H6 emphasized a ``scale threshold'': investment in complex visualization is justified only for research involving large-scale corpora, while for smaller datasets text and simple tables suffice.
Finally, publication media impose physical constraints, such as black-and-white printing, layout restrictions, and the absence of support for interactivity, that compel scholars to fall back on text.
H8 observed that the readability of high-density timelines deteriorates sharply in print, leading to a preference for combining narrative with simple tables.

These concerns and barriers collectively drive historians toward simple, disposable charts generated with general-purpose tools, rather than the complex visualization forms advocated by the visualization community.
These apprehensions stem not from technophobia but from methodological vigilance regarding the alignment of visualization with historical evidence and reasoning.
A profound understanding of these concerns is a prerequisite for domain-specific visualization innovations that are both usable and credible.

\section{Misuse and Problematic Visualization Patterns}

During coding (\cref{sec:coding}) and interviews (\cref{sec:interview}), historians and visualization researchers identified published figures in which design choices could hinder interpretation and may be considered as misuse.
We use ``misuse'' here in a narrow, practice-oriented sense: visualization choices that
(1) mismatch the structure of the underlying evidence or relationships,
(2) impede legibility and verification under typical publication constraints, or
(3) carry rhetorical force that exceeds what the underlying historical evidence can sustain.
These patterns can unintentionally shape readers' inferences in ways that conflict with historians' emphasis on accountable, revisitable claims (\cref{sec:motivations,sec:concerns}).

\begin{figure}
    \centering
    \includegraphics[width=1\linewidth]{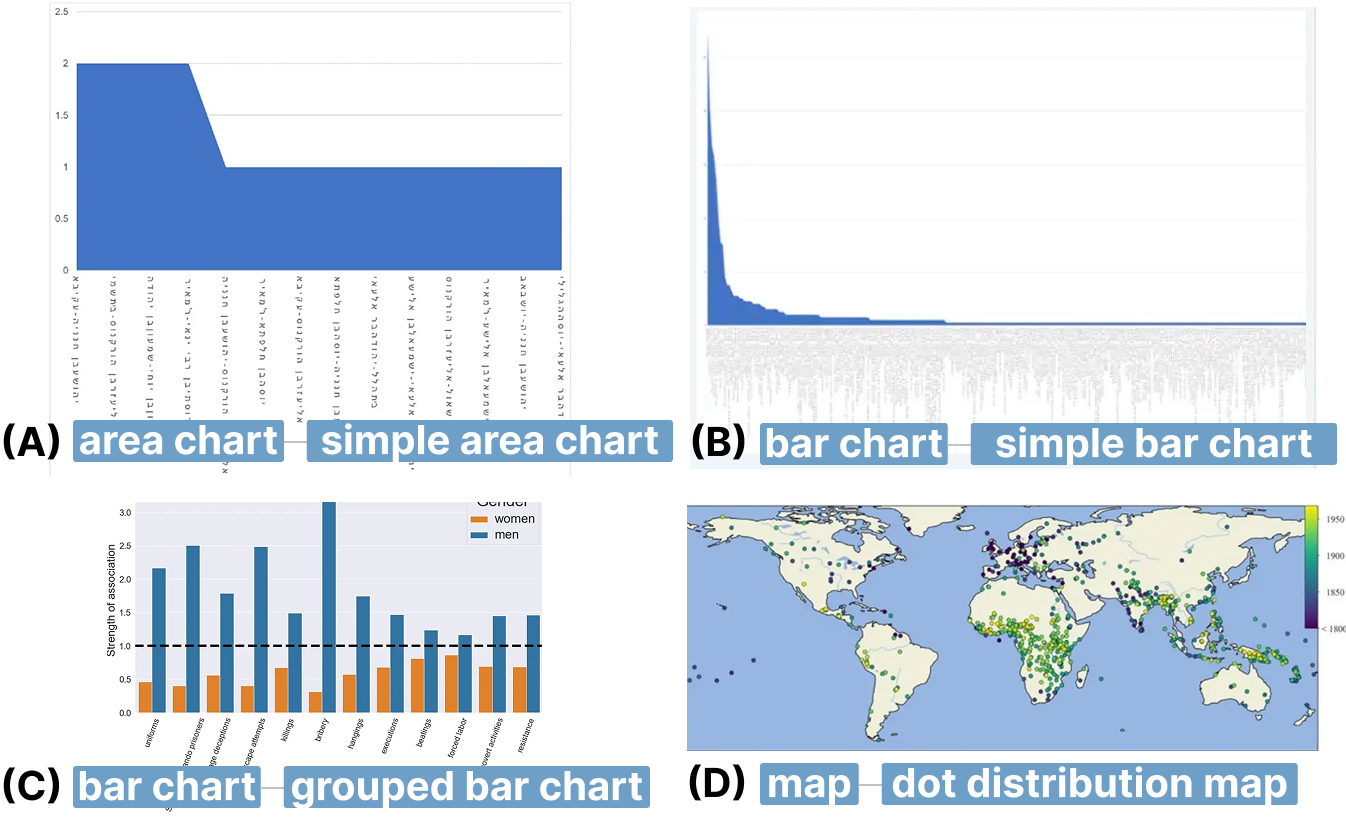}
    \caption{
        Representative examples of visualization misuse: 
        (A) Area chart applied to categorical data~\cite{Giovannetti2022Ontology}.
        (B) Illegible over-crowded category labels~\cite{Giovannetti2022Ontology}. 
        (C) Binary aggregation of gender identities~\cite{Toth2022Studying}. 
        (D) Temporal change encoded via color intensity~\cite{Kliger2022Translating}.
    }
    \label{fig:misuse}
\end{figure}

\textbf{Mismatch between form and data.}
Some misuses arise from selecting a chart form whose conventions imply relationships not supported by the data, compounded by presentation choices that prevent readers from checking the claim.
For instance, in a visualization depicting the ``Distribution of appearances of citation relationships in the Mishnat'' (\cref{fig:misuse}(A)), the author used an area chart for categorical counts~\cite{Giovannetti2022Ontology}.
Because area charts commonly cue continuity and smooth accumulation, this choice can suggest a trend structure that is not present.
Furthermore, its rotated category labels reduce legibility.
In another visualization showing the ``Distribution of appearances of complementary relationships in the Mishnat'' (\cref{fig:misuse}(B)), the category labels are so crowded that they become unreadable~\cite{Giovannetti2022Ontology}.

\textbf{Category oversimplification.}
A second pattern is collapsing historically contingent categories into overly coarse bins.
Binary gender bar charts were a frequent target: figures that aggregate complex identities into ``Male/Female'' columns (\cref{fig:misuse}(C))~\cite{Toth2022Studying}.
Interviewees emphasized that such aggregation can erase minorities in the historical record and produce what they described as ``overly simplified historical narratives.''

\textbf{Visual encodings impeding data readability.}
Misuse also appears when encodings make key historical distinctions hard to read or verify.
H10 critiqued a map that uses color intensity to represent temporal changes (\cref{fig:misuse}(D))~\cite{Kliger2022Translating}.
She described the mapping of time to color depth as ``ambiguous'', as the mapping makes it difficult to determine ``precisely when'' events occurred and to make fine-grained comparisons between regions.

H2, H7, and H10 also highlighted other potential visualization misuses in historical scholarship.
H2 noted problems such as the reliance on color encodings when publications are restricted to grayscale printing, the use of low-resolution screenshots, and readability issues caused by overlaying multiple visual encodings onto dense backgrounds or basemaps.
Similarly, H7 remarked that bar charts can become deceptive when axes are manipulated.
H10 further pointed out that using modern basemaps for historical inquiries runs the risk of implicitly framing ancient spaces through present-day boundaries, which may lead to anachronistic interpretations.

\section{Discussion}
\label{sec:discussion}

This section reflects on our mixed-methods approach, situates our taxonomy, and outlines implications and limitations.

\textbf{Methodological reflection.}
We propose a mixed-methods approach for interdisciplinary research that involves coding (\cref{sec:coding}) and interviews (\cref{sec:interview}).
The two methods served complementary roles.
The taxonomy and corpus labeling provided a quantitative map of visualization practice, preventing reliance on anecdotal examples and enabling systematic identification of underuse and misuse patterns that would be difficult to surface through interviews alone.
Conversely, \systemName as a boundary object (\cref{sec:system}) and the interviews grounded these patterns in historians' reasoning, revealing motivations, epistemological concerns, and contextual factors that corpus statistics cannot capture.
Our mixed-methods approach is potentially transferable for studying visualization usage in domains other than history.

\textbf{Comparison with existing taxonomies.}
Our taxonomy complements both scientific figure classifications (e.g., DocFigure~\cite{Jobin2019DocFigure}) and historical visualization corpora (e.g., VisTaxa~\cite{Zhang2025VisTaxa}).
Unlike work that classifies isolated images, we analyzed figures as they appear in complete research articles, preserving caption and textual context.
This contextual approach enabled us to identify the five functional roles of visualization in historical scholarship (\cref{sec:functions}), a dimension absent from existing image-focused taxonomies.
Compared to general visualization classification schemes (e.g., VIS30K~\cite{Chen2021VIS30K}), we provide finer divisions for maps and for chart types that frequently appear in historical writing.
Compared to VisTaxa~\cite{Zhang2025VisTaxa}, we additionally incorporate modern statistical chart categories (e.g., fitting curves).
A detailed comparison with existing classification schemes is provided in \iflabelexists{tab:glossary}{\cref{tab:glossary}}{the ``Glossary'' table in the supplemental material}.

\textbf{Future opportunities.}
Our findings point to several directions for future work.
First, historians' emphasis on provenance and source traceability (\cref{sec:motivations,sec:concerns}) suggests a need for visualization tools with built-in provenance tracking.
Second, the scarcity of temporal-structure visualizations despite time being central to historical inquiry (\cref{sec:taxonomy,sec:concerns}) motivates work on temporal representations that accommodate historical temporality while remaining legible to historians.
Third, the visualization literacy gap reported by both creators and audiences (\cref{sec:concerns}) calls for targeted training resources and tool designs that scaffold gradual adoption.
Finally, our interviews suggest that historians use specialized tools such as QGIS and Gephi more widely than publication records indicate.
Examining the tools historians use and the biases these tools may introduce constitutes a further research opportunity.

\textbf{Limitations.}
Our study has several limitations that qualify the scope of our claims.
First, our corpus and taxonomy are shaped by our decisions about what counts as a visualization and how to segment composite figures.
Second, the semi-automatic labeling pipeline scales the analysis but may introduce incorrect labels.
Although manual evaluation confirmed high top-level accuracy (89/100), readers should treat the reported distributions as approximate indicators rather than definitive measurements.
Finally, the corpus is limited to English-language research articles, excluding settings such as non-English scholarship and public history, each of which involves different audiences, incentives, and visualization traditions that warrant separate investigation.

\section{Conclusion}

We present a corpus-driven, mixed-methods study of how historians use visualization, combining analysis of \num{4831} visualization instances across \num{4142} research articles with semi-structured interviews with \num{11} historians.
Our findings identify five functional roles that visualizations serve in historical scholarship, document epistemological and practical barriers that impede adoption, and surface recurrent patterns of underuse and misuse.
These results provide a grounded foundation for designing visualization tools attuned to historians' evidentiary practices.

\section*{Acknowledgments}

We used GPT-5.1 solely for grammar correction and language polishing.
This work is supported by NSFC No.~62272012.
This work is partially supported by Wuhan East Lake High-Tech Development Zone National Comprehensive Experimental Base for Governance of Intelligent Society.

\bibliographystyle{eg-alpha-doi} 
\bibliography{assets/bibs/papers}
\fi

\ifshowappendix
\ifshowmain
\clearpage
\else

\fi
\appendix
\section{Article Subfields and Topics}
\label[appendix]{appendix:subfields}

\newcommand{\topicCategory}[1]{\textbf{#1}}

We assign each article in the corpus to one of eight history subfields.
The subfields are defined by grouping topics from the index of topics in \journalName{The American Historical Review}~\cite{TheAmericanHistoricalReview2019Index}.
The following lists the eight subfields and the topics assigned to each subfield from the index of topics~\cite{TheAmericanHistoricalReview2019Index}.
The topics are separated by a semicolon.

\begin{itemize}[leftmargin=*]
    \item \topicCategory{Legal \& Institutional History}:
    Administration; Institutions; Legal/Legislative; Reform
    
    \item \topicCategory{Economic History}:
    Agriculture; Business/Finance; Consumption/Consumers; Economic; Globalization; Industry; Trade

    \item \topicCategory{Military History}:
    Crime/Violence; Frontiers/Borderlands; Genocide; Military; Wars

    \item \topicCategory{Social History}:
    Careers/Professions; Childhood/Youth; Class; Diasporas; Elites; Emancipation; Ethnicity; Family; Gay/Lesbian; Gender; Health/Disease; Immigration/Migration; Indigenous Peoples; Labor; Local/Regional; Masculinity/Men; Medicine; Nobility; Peasants; Race/Racism; Sexuality; Slavery; Social History; Social Movements; Urban/Suburban; Women

    \item \topicCategory{Historiography}:
    Biography; Comparative; Historiography; Methods; Modernism/Modernity/Modernization; National Histories; Oral History; Public History; World

    \item \topicCategory{Intellectual History}:
    Anthropology/Archaeology; Education/Students; Ideology; Intellectual; Journalism; Language/Linguistics; Philosophy; Psychology/Psychiatry; Religion; Science/Technology; Theology; Theory

    \item \topicCategory{Cultural History}:
    Animals; Art/Architecture; Body; Cultural; Domesticity/Domestic; Emotions; Environment/Landscape; Exploration/Travel; Film/Photography; Food/Drink; Identity; Leisure/Entertainment; Literature; Material Culture; Media/Communications; Memory; Print/Manuscript Culture; Ritual/Celebration; Space/Place; Sports; Theater; Tourism

    \item \topicCategory{Political History}:
    Colonial/Postcolonial; Empire; Foreign Relations/Diplomatic; Human Rights/Humanitarianism; Nationalism; Peace; Political; Revolution; Rhetoric/Propaganda; State-Building/States; Terrorism/Espionage
\end{itemize}

\section{Prompt for Topic Extraction}
\label[appendix]{appendix:topic-prompt}

Based on the subfields categorized in \cref{appendix:subfields}, we designed the following system prompt to guide the model to extract the topic (i.e., subfield) for each paper.
For the user prompt, we concatenate the title and abstract of each article as the input text.
We manually verified the model's output on a random sample of 50 articles and found no misclassifications, suggesting the model can accurately assign subfields to articles based on the title and abstract.

\subsection{System Prompt}

You are an expert historian.
Your task is to (1) identify the most relevant topic label(s) from the controlled vocabulary below, then (2) map them to exactly one major category.
Return the result only as JSON that follows the required schema.

\textbf{Controlled Vocabulary}
\begin{itemize}
    \item Legal \& Institutional History: Administration; Institutions; Legal/Legislative; Reform;
    \item Economic History: Agriculture; Business/Finance; Consumption/Consumers; Economic; Globalization; Industry; Trade;
    \item Military History: Crime/Violence; Frontiers/Borderlands; Genocide; Military; Wars;
    \item Social History: Careers/Professions; Childhood/Youth; Class; Diasporas; Elites; Emancipation; Ethnicity; Family; Gay/Lesbian; Gender; Health/Disease; Immigration/Migration; Indigenous Peoples; Labor; Local/Regional; Masculinity/Men; Medicine; Nobility; Peasants; Race/Racism; Sexuality; Slavery; Social History; Social Movements; Urban/Suburban; Women;
    \item Historiography: Biography; Comparative; Historiography; Methods; Modernism/Modernity/Modernization; National Histories; Oral History; Public History; World;
    \item Intellectual History: Anthropology/Archaeology; Education/Students; Ideology; Intellectual; Journalism; Language/Linguistics; Philosophy; Psychology/Psychiatry; Religion; Science/Technology; Theology; Theory;
    \item Cultural History: Animals; Art/Architecture; Body; Cultural; Domesticity/Domestic; Emotions; Environment/Landscape; Exploration/Travel; Film/Photography; Food/Drink; Identity; Leisure/Entertainment; Literature; Material Culture; Media/Communications; Memory; Print/Manuscript Culture; Ritual/Celebration; Space/Place; Sports; Theater; Tourism;
    \item Political History: Colonial/Postcolonial; Empire; Foreign Relations/Diplomatic; Human Rights/Humanitarianism; Nationalism; Peace; Political; Revolution; Rhetoric/Propaganda; State-Building/States; Terrorism/Espionage.
\end{itemize}

\textbf{Instructions}

Read the input text and extract up to 3 most relevant topic labels exactly from the list above (string match).
Map those topics to one final category (choose the single best category reflecting the document's primary focus).
If nothing fits, choose the closest topic(s) and category by semantic proximity (still restricted to the vocabulary above).

Output JSON only, no extra text. Required JSON schema: 
\begin{verbatim}
{
  "topics": [
    {
      "label": "STRING (one of the topic labels above)",
      "category": "STRING (the category that contains this topic)",
      "confidence": 0.0
    }
  ],
  "final_topic": {
      "label": "STRING (one of: Legal & Institutional History, Economic History, Military History, Social History, Historiography, Intellectual History, Cultural History, Political History)",
      "confidence": 0.0
  }
}
\end{verbatim}

\section{Interview Procedure}
\label[appendix]{appendix:interview-procedure}

The following is the procedure of the interview described in \iflabelexists{sec:interview}{\cref{sec:interview}}{the ``Semi-Structured Interviews with Historians'' section in the main text}:

\begin{enumerate}[leftmargin=*]
    \item \textbf{Research background and prior experience.}
          Interviewees describe their research profile (specialization, typical data sources used in research, and research methods) and their prior experience working with visual materials and visualizations (e.g., maps, charts, diagrams, and digital tools).

    \item \textbf{Use, functions, and perceived value of visualization.}
          Interviewees are asked to reflect on how they use visualizations across stages of their work (e.g., exploring sources, organizing evidence, communicating arguments) and to discuss concrete examples.
          The interviewer probes for intended functions, perceived benefits, and practical constraints or worries that shape decisions about whether and how to visualize.

    \item \textbf{Corpus and taxonomy-based exploration with \systemName.}
          After a brief tutorial, interviewees use \systemName to browse figures across journals and subfields, inspect taxonomy labels and summary statistics, and reflect on whether observed patterns matched or contradicted their expectations and discuss potential reasons behind the patterns.
          They are encouraged to search for topics, authors, or keywords relevant to their expertise and to comment on visualization usage in related articles.

    \item \textbf{Avoided visualizations, risks, and concerns.}
          Finally, interviewees are asked to identify visualization types or design choices they tend to avoid or distrust and why, and to articulate perceived risks.
\end{enumerate}

\section{Visualization Taxonomy}
\label[appendix]{appendix:visualization-taxonomy}

\Cref{fig:taxonomy} summarizes the visualization taxonomy developed by coding the corpus (see \iflabelexists{sec:coding}{\cref{sec:coding}}{the ``Coding Process and Taxonomy Construction'' subsection of the main paper}).

\begin{figure}
    \centering
    \includegraphics[width=\linewidth]{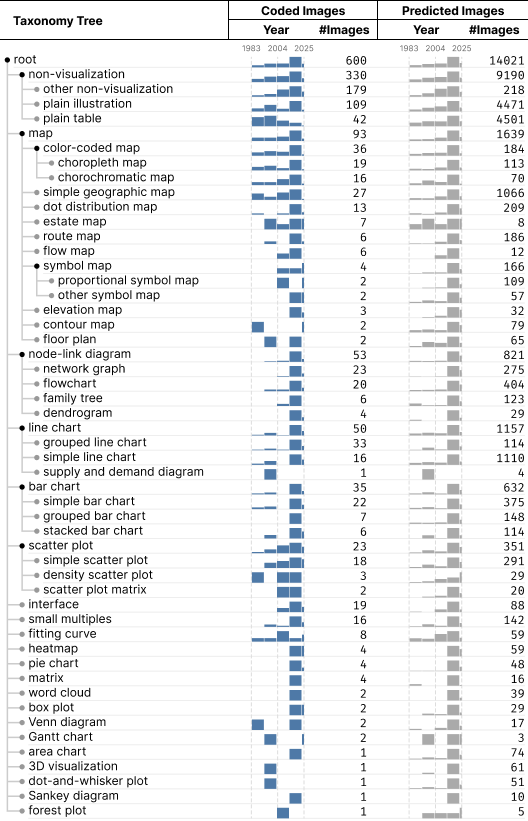}
    \caption{%
        Visualization taxonomy developed by coding the corpus.
        The taxonomy contains \nTaxaFinalAllVis visualization-related taxa (excluding the \taxon{root}, \taxon{non-visualization}, and the subcategories of \taxon{non-visualization}), among which \nTaxaFinalFirstLevelVis are at the first level below the root and \nTaxaFinalLeafVis are leaves.
        In the ``Coded Images'' column, ``\#Images'' reports how many images were assigned to each taxon by the coders, and ``Year'' column summarizes the publication years of those figures.
        An image may be labeled with multiple taxa.
        Model-predicted labels on the full corpus shown in the ``Predicted Images'' column are imperfect and should be interpreted cautiously.
    }
    \label{fig:taxonomy}
\end{figure}

\section{Taxon Definitions}
\label[appendix]{appendix:taxon-definitions}

\Cref{fig:examples} shows example images belonging to each leaf taxon in the taxonomy tree developed through the iterative coding process described in \iflabelexists{sec:coding}{\cref{sec:coding}}{the ``Coding Process and Taxonomy Construction'' subsection of the main paper}.
\Cref{tab:taxon-definitions} lists the definition for each taxon.
These definitions are intended as working scaffolds rather than definitive characterizations.

\begin{figure*}[!htbp]
    \centering
    \includegraphics[width=0.9\linewidth]{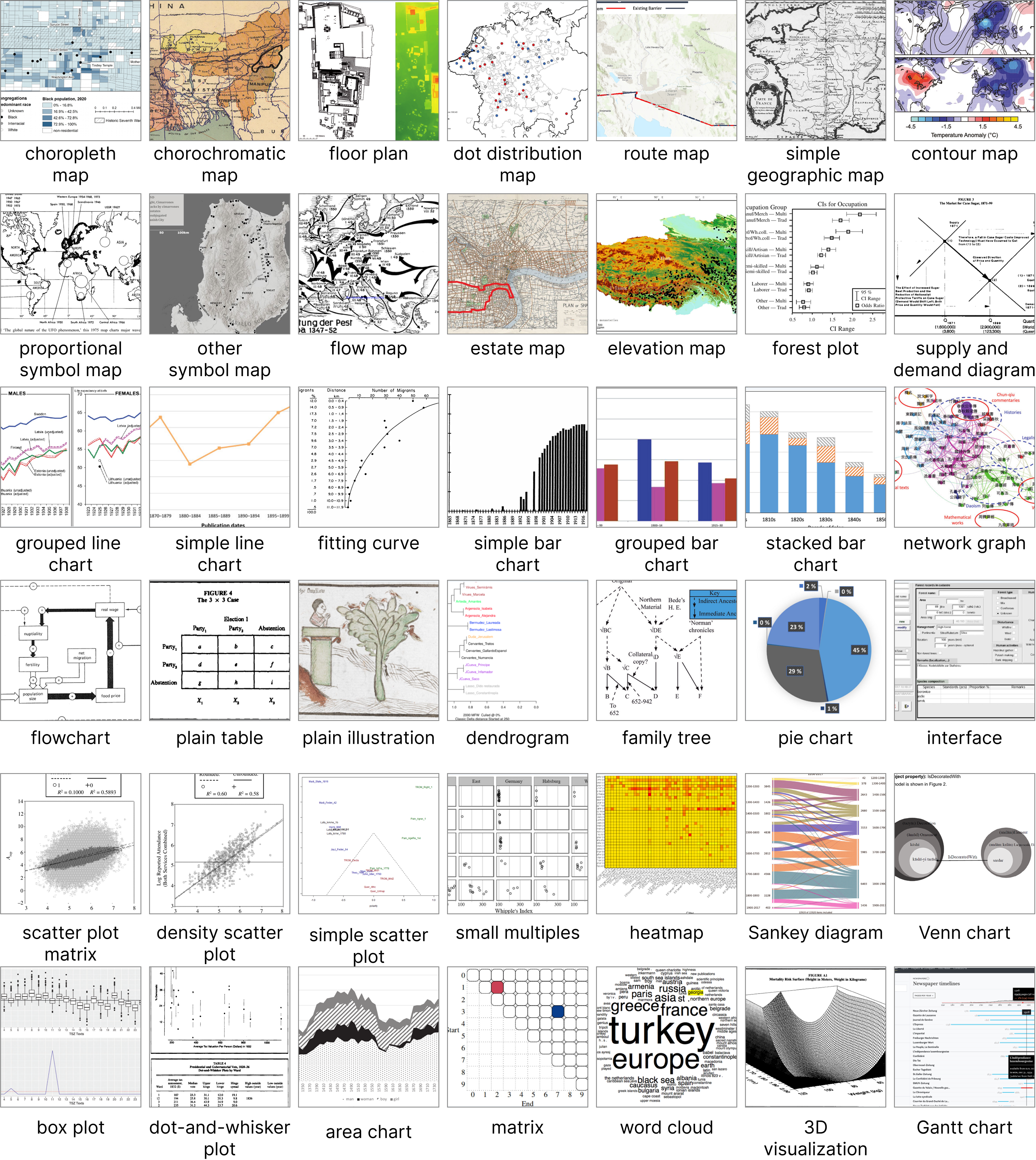}
    \caption{
        \textbf{Example images in each leaf taxon:}
        Among the \nTaxaFinalAll taxa, excluding \taxon{other non-visualization}, there are 42 leaf nodes in the taxonomy tree.
        We show an example image of each leaf node in this image matrix.
        Note that some images may have been assigned to more than one leaf node by the coders.
        Some images are cropped to show representative design of specific taxon more clearly.
        The definition of each leaf taxon can be found in \cref{tab:taxon-definitions}.
    }
    \label{fig:examples}
\end{figure*}

\let\lw\relax
\newlength\lw
\setlength\lw{\dimexpr .35\textwidth - 2\tabcolsep}
\newlength\rw
\setlength\rw{\dimexpr .65\textwidth - 2\tabcolsep}
\newcommand{\leafTaxon}[1]{#1}
\newcommand{\hypAnc}[1]{\ensuremath{\sqrt{\mathrm{#1}}}}

\begin{table*}[!htbp]
    \caption{
    \textbf{Taxon definitions:}
    This table provides the definition of each taxon.
    }
    \label{tab:taxon-definitions}
    \centering
    \scriptsize
    \begin{tabular}{p{\lw}p{\rw}}
        \toprule
        \textbf{Taxon} & \textbf{Definition} \\
        \midrule
        visualization & A [visual representation] commonly uses [visual design] to represent [data]. \\
        \midrule
        map & A [visual representation] commonly uses [visual symbols and colors] to represent [geographical information like areas, objects, and themes with their spatial relationships and features]. \\
        \midrule
        map > \leafTaxon{simple geographic map} & A [map] commonly uses [outlines] to represent [basic geographic features and boundaries], [without thematic or statistical overlays]. Example: \href{https://doi.org/10.2307/2162692}{``A simple map of France showing its provinces and neighboring regions in the 18th century.''} \\
        \midrule
        map > \leafTaxon{flow map} & A [map] commonly uses [lines or arrows] to represent [non-physical flows such as trade or the movement of objects between different areas]. Example: \href{https://doi.org/10.1093/pastj/gtq069}{``A map showing the spread of the Black Death in Europe from 1347 to 1352 with arrows indicating routes of contagion.''} \\
        \midrule
        map > \leafTaxon{estate map} & A [map] commonly uses [plats] to represent [extensive landed property, estate boundaries and details such as buildings and streets, usually in a city or town]. Example: \href{https://doi.org/10.1093/llc/fqad044}{``A map of Shanghai highlighting the central urban districts west of the Huangpu River, with a red boundary marking a specific study area.''} \\
        \midrule
        map > color-coded map & A [map] commonly uses [color encoding] to represent [different data features between areas based on certain criteria such as climates, populations, or land use]. \\
        \midrule
        map > color-coded map > \leafTaxon{choropleth map} & A [map] commonly uses [colored, shaded, or patterned areas] to represent [statistical variables proportionally across pre-defined geographic areas, with each area filled with a uniform color, shade, or pattern]. Example: \href{https://doi.org/000609.xml}{``A map showing Black population distribution and congregation locations in Philadelphia's historic Seventh Ward in 2020.''} \\
        \midrule
        map > color-coded map > \leafTaxon{chorochromatic map} & A [map] commonly uses [color encoding] to represent [non-overlapping, categorical data-driven areas of different categories without relying on pre-existing boundaries]. [The areas can be discrete]. Example: \href{https://doi.org/10.1093/pastj/gty045}{``A map of Northeast India and surrounding regions, highlighting Assam, Manipur, Tripura, Bhutan, East Pakistan, and Burma, with a black boundary marking a specific area of interest.''} \\
        \midrule
        map > \leafTaxon{elevation map} & A [map] commonly uses [numbers or scales] to represent [the height of elevations and terrain]. Example: \href{https://doi.org/10.1093/llc/fqad103}{``A map of the Tibetan Plateau showing elevation and the distribution of Tibetan Buddhist monasteries across Ü-Tsang, Kham, and Amdo.''} \\
        \midrule
        map > \leafTaxon{dot distribution map} & A [map] commonly uses [dots, usually circles] to represent [the distribution and density of a specified quantity across a pre-defined geographic area]. Example: \href{https://doi.org/10.1080/01615440.2015.1033582}{``A map of the Holy Roman Empire circa 1050 showing guild participation, with red dots for major forces, blue for participation, and grey for none.''} \\
        \midrule
        map > \leafTaxon{contour map} & A [map] commonly uses [contour lines] to represent [continuous data distribution in a two-dimensional area], with [each line connecting points of equal value for a specific measurement]. Example: \href{https://doi.org/10.2307/1862554}{``A map of the Gulf of Mexico region showing Native American tribes, Spanish settlements, and French outposts in the late 17th century.''} \\
        \midrule
        map > symbol map & A [map] commonly uses [symbols] to represent [various features or data points about specific locations or areas]. \\        \midrule
        map > symbol map > \leafTaxon{other symbol map} & A [map] commonly uses [unproportional symbols] to represent [various types of information associated with different areas or locations]. Example: \href{https://doi.org/10.1093/ahr/rhad298}{``A map of northern Luzon, Philippines, showing Spanish colonial cities, Cimarrones' flight routes and attacks, apostates, and unsubjugated mountain areas.''} \\
        \midrule
        map > symbol map > \leafTaxon{propotional symbol map} & A [map] commonly uses [proportional symbols of varying sizes] to represent [the magnitude of quantitative variables associated with different areas or locations, such as quantitative data]. Example: \href{https://doi.org/10.1080/07341512.2012.723340}{``A world map charting major waves of UFO sightings from 1947 to the 1970s, with circles marking events in North America, Europe, Asia, Africa, and South America.''} \\
        \midrule
        map > \leafTaxon{route map} & A [map] commonly uses [lines and symbols] to represent [physical routes such as roads, paths, and streets, focusing on human-made roads or transport links rather than natural geographical information]. Example: \href{https://doi.org/000626.xml}{``A map of the U.S.-Mexico border in California and Arizona, showing new border wall segments in red and existing barriers in blue.''} \\
        \midrule
        map > \leafTaxon{floor plan} & A [map] commonly uses [scaled drawings with walls, rooms, and structural elements] to represent [the layout of a building's interior, including spaces, functions, and circulation]. Example: \href{https://doi.org/10.1093/llc/fqx014}{``A floor plan and heatmap of an archaeological site, showing detailed building layouts alongside elevation-based spatial visualization.''} \\
        \midrule
        line chart & A [visual representation] commonly uses [positions along the x and y axes with straight or curved line segments connecting data points] to represent [changes in values over time or across categories]. \\
        \midrule
        line chart > \leafTaxon{grouped line chart} & A [line chart] commonly uses [multiple lines in the same coordinate system] to represent [different groups or categories for comparison over the same time span or x-axis values]. Example: \href{https://doi.org/10.1080/01615440.2017.1338977}{``A pair of line charts comparing male and female life expectancy at birth in Sweden, Finland, Latvia, Estonia, and Lithuania from 1923 to 1938, with adjusted and unadjusted data series.''} \\
        \midrule
        line chart > \leafTaxon{simple line chart} & A [line chart] commonly uses [a single continuous line] to represent [changes of one variable over time or across categories]. Example: \href{https://doi.org/10.1093/llc/fqaa005}{``A line chart showing the average number of characters prone to hysteria per play across different publication periods from before 1869 to after 1905, with a marked increase after 1895.''} \\
        \midrule
        line chart > \leafTaxon{fitting curve} & A [line chart] commonly uses [a smooth curve fitted to data points] to represent [the underlying trend or functional relationship in the dataset]. Example: \href{https://doi.org/10.2307/1859923}{``A scatter plot with a fitted curve showing the distance decay of migration, where most moves occur within 4 km and long-distance moves beyond 8 km are rare.''} \\
        \midrule
        line chart > \leafTaxon{supply and demand diagram} & A [supply and demand diagram] commonly uses [two intersecting curves plotted on price-quantity axes] to represent [how market equilibrium changes]. Example: \href{https://doi.org/10.1080/01615440109600793}{``A supply and demand diagram showing the world market for cane sugar between 1871 and 1899, where prices fell from 5.4 to 2.5 cents per pound while production rose from 1.6 to 2.9 million tons.''} \\
        \midrule
        bar chart & A [visual representation] commonly uses [rectangular bars with lengths proportional to the corresponding values] to represent [categorical and quantitative data]. \\
        \midrule
        bar chart > \leafTaxon{simple bar chart} & A [bar chart] commonly uses [simple rectangular bars with lengths proportional to the values] to represent [quantitative data] without [additional decoration or other statistical characteristics]. Example: \href{https://doi.org/10.2307/2168774}{``A bar chart showing the number of cooperatives reporting from 1865 to 1925, with rapid growth after 1895, stagnation around 1910--1919, and a sharp rise to nearly 6000 by 1925.''} \\
        \midrule
        bar chart > \leafTaxon{stacked bar chart} & A [bar chart] commonly uses [stacked bars on top of each other] to represent [the total and segment proportions of different groups within the same category for cumulative data]. Example: \href{https://doi.org/10.1093/llc/fqac088}{``A stacked bar chart showing the decline in sentence lengths from the 1790s to the 1870s, with 10--14 word sentences dominating but gradually decreasing, and long sentences of 20+ words nearly disappearing by the mid-19th century.''} \\
        \midrule
        bar chart > \leafTaxon{grouped bar chart} & A [bar chart] commonly uses [multiple bars for each categorical group placed side by side] to represent [different data points to facilitate comparison between categories]. Example: \href{https://doi.org/10.1080/01615440.2013.821876}{``A grouped bar chart comparing White, Black, and Unknown populations across 1885--99, 1900--14, and 1915--30, showing a decline in White numbers, a peak in Black numbers around 1900--14, and relatively stable but lower counts for Unknown.''} \\
        \midrule
        node-link diagram & A [visual representation] commonly uses [nodes and links] to represent [networks of entities and their relationships or flows], [typically with at least a two-tier node-link structure]. \\
        \bottomrule
    \end{tabular}
\end{table*}

\begin{table*}[!htbp]
    \ContinuedFloat
    \caption{continued}
    \centering
    \scriptsize
    \begin{tabular}{p{\lw}p{\rw}}
        \toprule
        \textbf{Taxon} & \textbf{Definition} \\
        \midrule
        node-link diagram > \leafTaxon{network graph} & A [visual representation] commonly uses [nodes and links] to represent [networks of entities and their relationships or flows], [typically with at least a two-tier node-link structure]. Example: \href{https://doi.org/10.1093/llc/fqx024}{``A network graph mapping ancient Chinese texts by genre, with Chun-qiu commentaries at the core, Legalist and warfare works clustered together, and medical and mathematical texts more peripheral.''} \\
        \midrule
        node-link diagram > \leafTaxon{flowchart} & A [node-link diagram] commonly uses [ordered stages linked by lines, arrows, or flowing bands] to represent [a process, workflow, decision sequence, or the progression of entities, materials, or information across sequential stages]. Example: \href{https://doi.org/10.1080/07341512.2021.1891394}{``A flowchart showing causal loops among population, fertility, mortality, wages, and food prices, illustrating a Malthusian economic-demographic model.''} \\
        \midrule
        node-link diagram > \leafTaxon{dendrogram} & A [node-link diagram] commonly uses [hierarchically branching nodes and links] to represent [tree-structured data such as taxonomies or clustering results]. Example: \href{https://doi.org/000346.xml}{``A dendrogram clustering Spanish literary works by most frequent words, showing clear groupings of Virues, Argensola, Bermúdez, Cervantes, and J. Cueva.''} \\
        \midrule
        node-link diagram > \leafTaxon{family tree} & A [node-link diagram] commonly uses [family members as nodes and kinship relations as links] to represent [family structures and hereditary patterns across generations]. Example: \href{https://doi.org/10.1093/llc/fqw014}{``A family tree showing the transmission of chronicles from an original source through hypothetical ancestors (\checkmark A, \checkmark BC, \checkmark DE) to extant versions A--F, with both immediate and indirect ancestral links.''} \\
        \midrule
        \leafTaxon{interface} & A [visual representation] commonly uses [interactive layouts, timelines, dashboards, or navigable views] to represent [an intermediate stage between simple demonstration and in-depth research, allowing the audience to explore data, manipulate views, and access additional layers of information]. Example: \href{https://doi.org/10.1080/01615440.2018.1444523}{``An interface for recording and managing cadastre-based forest data, including forest area, management type, disturbances, human activities, and species composition.''} \\
        \midrule
        scatter plot & A [visual representation] commonly uses [points positioned along x and y axes] to represent [the relationship between two quantitative variables]. \\
        \midrule
        scatter plot > \leafTaxon{scatter plot matrix} & A [scatter plot] commonly uses [a grid of scatter plots arranged in a matrix] to represent [pairwise relationships among multiple variables]. Example: \href{https://doi.org/10.3200/HMTS.39.1.24-46}{``A scatter plot matrix showing the relationship between Vex and Arep, with two regression fits (R² = 0.10 and R² = 0.59).''} \\
        \midrule
        scatter plot > \leafTaxon{density scatter plot} & A [scatter plot] commonly uses [point shading, transparency, or color gradients] to represent [the density of overlapping data points in crowded regions]. Example: \href{https://doi.org/10.3200/HMTS.39.1.24-46}{``A scatter plot showing church seating capacity versus attendance, with regression lines indicating a strong positive correlation ($R^2 \approx 0.58-0.60$).''} \\
        \midrule
        scatter plot > \leafTaxon{simple scatter plot} & A [scatter plot] commonly uses [single points in a two-dimensional plane] to represent [the distribution and possible correlation between two variables]. Example: \href{https://doi.org/000371.xml}{``A scatter plot mapping textual sources by polarity and credence score, with a triangular boundary highlighting negative, neutral, and positive cases.''} \\
        \midrule
        \leafTaxon{small multiples} & A [visual representation] commonly uses [a series of repeated charts with consistent axes and scales] to represent [comparisons across categories, time periods, or conditions, usually with more than three repeated charts]. Example: \href{https://doi.org/10.1080/01615440.2017.1393359}{``A small multiples scatter plot matrix showing Whipple's Index across Balkans, East, Germany, Habsburg, and West, divided into five value ranges.''} \\
        \midrule
        \leafTaxon{heatmap} & A [visual representation] commonly uses [color encoding in a grid or matrix] to represent [the magnitude of values across two dimensions]. Example: \href{https://doi.org/10.1093/llc/fqae037}{``A heatmap showing citation relationships, with rows as cited texts, columns as citing texts, and color intensity indicating citation frequency.''} \\
        \midrule
        \leafTaxon{Sankey diagram} & A [visual representation] commonly uses [flows with varying thickness] to represent [quantities of flow or transfer between entities in a network]. Example: \href{https://doi.org/10.1093/llc/fqz058}{``A Sankey diagram showing the flow of 22,920 texts from their original composition periods (1200--2017) to later periods of reuse, with thicker bands indicating more works being transmitted across centuries.''} \\
        \midrule
        \leafTaxon{Venn chart} & A [visual representation] commonly uses [overlapping circles or other shapes] to represent [logical relationships, similarities, and differences between sets]. Example: \href{https://doi.org/000559.xml}{``A Venn chart illustrating overlapping categories of architectural elements and decorations, showing how carved tile (kāshī-yi tarāshī) connects to larger structural components (sardar) through shared attributes.''} \\
        \midrule
        \leafTaxon{box plot} & A [visual representation] commonly uses [rectangles with quartile lines and whiskers] to represent [the distribution, spread, and outliers of a dataset]. Example: \href{https://doi.org/10.1093/llc/fqy055}{``A boxplot and line chart comparing 28 TSZ texts, showing distribution differences in MuSi and Ritter metrics with most p-values near zero except peaks at texts 5 and 11.''} \\
        \midrule
        \leafTaxon{dot-and-whisker plot} & A [visual representation] commonly uses [dots with horizontal or vertical whiskers] to represent [point estimates and their associated confidence intervals or uncertainty]. Example: \href{https://doi.org/10.1080/01615449809600090}{``A dot-and-whisker plot with table showing links between 1832 tax valuation per person and Democratic vote share (1828--1836) across wards, highlighting variations and outliers.''} \\
        \midrule
        \leafTaxon{matrix} & A [visual representation] commonly uses [a grid of rows and columns] to represent [relationships or values between two categorical dimensions]. Example: \href{https://doi.org/10.1093/llc/fqab032}{``A matrix diagram showing start-to-end transitions, with a red cell at (1,2) and a blue cell at (3,7).''} \\
        \midrule
        \leafTaxon{word cloud} & A [visual representation] commonly uses [words sized proportionally to their frequency or importance] to represent [the prominence of terms within a text corpus]. Example: \href{https://doi.org/000250.xml}{``A word cloud highlighting geographic and historical terms, with `turkey' and `europe' as the most prominent words.''} \\
        \midrule
        \leafTaxon{3D visualization} & A [visual representation] commonly uses [three-dimensional coordinate systems, surfaces, meshes, or volumetric forms] to represent [multivariate relationships or spatial structures that are difficult to express on a 2D plane]. Example: \href{https://doi.org/10.1080/01615440.1993.10594215}{``A 3D visualization showing mortality risk by height and weight, lowest at moderate levels and higher at underweight or overweight extremes.''} \\
        \midrule
        \leafTaxon{Gannt chart} & A [visual representation] commonly uses [horizontal bars along a timeline] to represent [tasks, their duration, and dependencies in project management]. Example: \href{https://doi.org/10.1080/01615440.2024.2344004}{``A Gantt chart showing newspaper publication timelines, with bars for active years, gaps for missing issues, and markers for key dates like 1908.''} \\
        \midrule
        \leafTaxon{area chart} & A [visual representation] commonly uses [filled areas between lines and a baseline] to represent [quantities such as cumulative data or multiple data series over time]. Example: \href{https://doi.org/10.1093/pastj/gtaa005}{``A stacked area chart showing word frequencies of `man', `woman', `boy', and `girl' from 1270 to 1830, with `man' dominating and the others fluctuating over time.''} \\
        \midrule
        pie chart & A [visual representation] commonly uses [a divided circle or donut into slices] to represent [numerical proportions of data segments]. \\
        \midrule
        \leafTaxon{forest chart} & A [visual representation] commonly uses [horizontal lines with square markers and confidence intervals] to represent [the results of multiple statistical studies, typically in a meta-analysis]. Example: \href{https://doi.org/10.3200/HMTS.37.1.5-22}{``A forest plot showing odds ratios and 95\% confidence intervals for different occupations, with Manufacturing/Merchant highest (above 2.0) and most other groups clustering around 1.0--1.5.''} \\
        \midrule
        non-visualization & An [image] that [does not qualify as a visualization]. [This category is used to mark images that fall outside the scope of a visualization taxonomy]. \\
        \midrule
        non-visualization > \leafTaxon{plain table} & A [non-visualization] commonly uses [rows and columns of cells] to represent [structured data]. [Typically, more than two rows and columns are used]. Example: \href{https://doi.org/000382.xml}{``A 3×3 contingency table showing voter shifts between Party\textsubscript{1}, Party\textsubscript{2}, and Abstention across two elections.''} \\
        \midrule
        non-visualization > \leafTaxon{plain illustration} & A [non-visualization] commonly uses [drawings, sketches, or paintings] to represent [visual explanations of concepts, plans, processes, or scenes]. Example: \href{https://doi.org/10.1080/01615440.1991.9955296}{``A medieval illustration labeled `perching of the swallow', showing a person at a podium, swallows perched on a leafy plant, and a hand holding a bird on the right.''} \\
        \bottomrule
    \end{tabular}
\end{table*}

\begin{table*}[ht]
    \caption{
        \textbf{Glossary:}
        Terms collected from taxonomy-related papers.
        The ``Terms Extracted From'' column indicates where each paper lists or illustrates its category terms.
        The glossary from the related papers is used for coders' reference in the coding process.
        The final row shows our taxonomy terms for comparison.
    }
    \label{tab:glossary}
    \centering
    \scriptsize
    \renewcommand{\arraystretch}{1}
    \begin{tabular}{c c c c c}
        \toprule
        \textbf{ID} & \textbf{Paper}                              & \textbf{Venue} & \textbf{Year} & \textbf{Terms Extracted From}                                                                                                                                                                                                                                                                                                                                                                                                                                                                                                                                                                                                                                                                                                                                                                                                                                                                                                                                                                                                                                        \\
        \midrule
        1           & Jobin2019DocFigure\cite{Jobin2019DocFigure} & ICDAR Workshop & 2019          & Figure 1                                                                                                                                                                                                                                                                                                                                                                                                                                                                                                                                                                                                                                                                                                                                                                                                                                                                                                                                                                                                                                                             \\
        \midrule
        \multicolumn{5}{p{0.95\linewidth}}{\raggedright\textbf{Terms:} ``Line graph'', ``Natural image'', ``Table'', ``3Dobject'', ``Bar plot'', ``Scatter plot'', ``Medical image'', ``Sketch'', ``Geographic map'', ``Flowchart'', ``Heat map'', ``Mask'', ``Block diagram'', ``Venn diagram'', ``Confusion matrix'', ``Histogram'', ``Box plot'', ``Vector plot'', ``Pie chart'', ``Surface plot'', ``Algorithm'', ``Contour plot'', ``Tree diagram'', ``Bubble chart'', ``Polar plot'', ``Area chart'', ``Pareto chart'', ``Radar chart''}                                                                                                                                                                                                                                                                                                                                                                                                                                                                                                                                                                                                                            \\
        \toprule
        2           & Chen2021VIS30K\cite{Chen2021VIS30K}         & TVCG           & 2021          & TABLE 1                                                                                                                                                                                                                                                                                                                                                                                                                                                                                                                                                                                                                                                                                                                                                                                                                                                                                                                                                                                                                                                              \\
        \midrule
        \multicolumn{5}{p{0.95\linewidth}}{\raggedright\textbf{Terms:} ``Table'', ``Matrix and parallel coordinates'', ``Area and circles'', ``Multiple types'', ``Bars'', ``Photos'', ``Bullets and equations'', ``Point-based'', ``Line chart'', ``Scientific data visualization'', ``Maps'', ``Tree and Networks''}                                                                                                                                                                                                                                                                                                                                                                                                                                                                                                                                                                                                                                                                                                                                                                                                                                                    \\
        \toprule
        3           & Karishma2023ACL\cite{Karishma2023ACLFig}    & AAAI workshop  & 2023          & Figure 1                                                                                                                                                                                                                                                                                                                                                                                                                                                                                                                                                                                                                                                                                                                                                                                                                                                                                                                                                                                                                                                             \\
        \midrule
        \multicolumn{5}{p{0.95\linewidth}}{\raggedright\textbf{Terms:} ``Algorithms'', ``Architecture Diagram'', ``Bar Charts'', ``Box Plot'', ``Confusion Matrix'', ``Graph'', ``Line graph chart'', ``Geographical Maps'', ``Natural images'', ``Neural networks'', ``NLP text grammar'', ``Pareto'', ``Pie chart'', ``Scatter plot'', ``Screenshots'', ``Tables'', ``Tree'', ``Venn diagram'', ``Word cloud''}                                                                                                                                                                                                                                                                                                                                                                                                                                                                                                                                                                                                                                                                                                                                                         \\
        \toprule
        4           & Deng2023VisImages\cite{Deng2023VisImages}   & TVCG           & 2023          & TABLE 3                                                                                                                                                                                                                                                                                                                                                                                                                                                                                                                                                                                                                                                                                                                                                                                                                                                                                                                                                                                                                                                              \\
        \midrule
        \multicolumn{5}{p{0.95\linewidth}}{\raggedright\textbf{Terms:} ``area'', ``area chart'', ``proportional area chart (PAC)'', ``bar'', ``bar chart'', ``circle'', ``donut chart'', ``pie chart'', ``diagram'', ``flow diagram'', ``chord diagram'', ``sankey diagram'', ``venn diagram'', ``statistic'', ``box plot'', ``error bar'', ``stripe graph'', ``table'', ``line'', ``contour graph'', ``line chart'', ``storyline'', ``polar plot'', ``parallel coordinate (PCP)'', ``surface graph'', ``vector graph'', ``map'', ``point'', ``scatter plot'', ``grid matrix'', ``heatmap'', ``matrix'', ``text'', ``phrase net'', ``word cloud'', ``word tree'', ``graph tree'', ``graph'', ``tree'', ``treemap'', ``hierarchical edge bundling (HEB)'', ``sunburst/icicle plot'', ``special'', ``glyph-based visualization'', ``unit visualization''}                                                                                                                                                                                                                                                                                                                   \\
        \toprule
        5           & Zhang2025VisTaxa\cite{Zhang2025VisTaxa}     & TVCG           & 2025          & Figure 4                                                                                                                                                                                                                                                                                                                                                                                                                                                                                                                                                                                                                                                                                                                                                                                                                                                                                                                                                                                                                                                             \\
        \midrule
        \multicolumn{5}{p{0.95\linewidth}}{\raggedright\textbf{Terms:} ``map'', ``color-coded map'', ``choropleth map'', ``chorochromatic map'', ``dasymetric map'', ``ther color-coded map'', ``route map'', ``bathymetric map'', ``contour map'', ``symbol map'', ``proportional symbol map'', ``other symbol map'', ``flow map'', ``estate map'', ``elevation map'', ``dot distribution map'', ``simple geographic map'', ``pictorial map'', ``symbolic map'', ``bar chart'', ``grouped bar chart'', ``simple bar chart'', ``pictogram bar chart'', ``stacked bar chart'', ``diverging bar chart'', ``bent bar chart'', ``table'', ``illustration'', ``non-visualization'', ``plain illustration'', ``plain text'', ``plain map'', ``plain table'', ``ISOTYPE'', ``line char'', ``pie chart'', ``glyph'', ``unit chart'', ``celestial map'', ``star map'', ``solar system map'', ``sun map'', ``lunar map'', ``area chart'', ``simple area chart'', ``step area chart'', ``node-link diagram'', ``Sankey diagram'', ``timeline'', ``wind rose'', ``other statistical chart'', ``rose chart'', ``parallel coordinates'', ``radar chart'', ``treemap'', ``violin plot''} \\
        \toprule
        6           & Our work                                    & --             & --            & --                                                                                                                                                                                                                                                                                                                                                                                                                                                                                                                                                                                                                                                                                                                                                                                                                                                                                                                                                                                                                                                                   \\
        \midrule
        \multicolumn{5}{p{0.95\linewidth}}{\raggedright\textbf{Terms:} ``non-visualization'', ``other non-visualization'', ``plain illustration'', ``plain table'', ``map'', ``color-coded map'', ``choropleth map'', ``chorochromatic map'', ``simple geographic map'', ``dot distribution map'', ``estate map'', ``flow map'', ``route map'', ``symbol map'', ``proportional symbol map'', ``other symbol map'', ``elevation map'', ``floor plan'', ``contour map'', ``line chart'', ``grouped line chart'', ``simple line chart'', ``fitting curve'', ``supply and demand diagram'', ``node-link diagram'', ``network graph'', ``flowchart'', ``family tree'', ``dendrogram'', ``bar chart'', ``simple bar chart'', ``grouped bar chart'', ``stacked bar chart'', ``scatter plot'', ``simple scatter plot'', ``density scatter plot'', ``scatter plot matrix'', ``interface'', ``small multiples'', ``heatmap'', ``matrix'', ``pie chart'', ``box plot'', ``venn chart'', ``gantt chart'', ``word cloud'', ``dot-and-whisker plot'', ``area chart'', ``sankey diagram'', ``3D visualization'', ``forest plot''}                                                        \\
        \bottomrule
    \end{tabular}
\end{table*}

\ifshowmain\else
\bibliographystyle{eg-alpha-doi}
\bibliography{assets/bibs/papers.bib}

\newcommand{\etalchar}[1]{$^{#1}$}
\begin{thebibliography}{\uppercase{LVMVA{\etalchar{*}}21}}

\bibitem[ABB{\etalchar{*}}26]{Almas2026Historical}
\textsc{Alm{\aa}s I., Berger T., Boppart T., Burchardi K., Ejermo O., Eriksson
  B., Larsson A., Malmberg H., Maukner S., Olsson M., Ostermeyer V.}:
\newblock Historical manufacturing census of {Sweden}: Data description and
  quality assessment.
\newblock \emph{Historical Methods: A Journal of Quantitative and
  Interdisciplinary History 59}, 1 (2026), 20--38.
\newblock \href {https://doi.org/10.1080/01615440.2025.2527132}
  {\path{doi:10.1080/01615440.2025.2527132}}.

\bibitem[ALC{\etalchar{*}}23]{Akbaba2023Troubling}
\textsc{Akbaba D., Lange D., Correll M., Lex A., Meyer M.}:
\newblock Troubling collaboration: Matters of care for visualization design
  study.
\newblock In \emph{Proceedings of the CHI Conference on Human Factors in
  Computing Systems} (2023), ACM.
\newblock \href {https://doi.org/10.1145/3544548.3581168}
  {\path{doi:10.1145/3544548.3581168}}.

\bibitem[BC21]{Braun2021Thematic}
\textsc{Braun V., Clarke V.}:
\newblock \emph{Thematic Analysis: A Practical Guide}.
\newblock SAGE Publications, 2021.

\bibitem[BCC{\etalchar{*}}25]{Bai2025Qwen3VL}
\textsc{Bai S., Cai Y., Chen R., Chen K., Chen X., Cheng Z., Deng L., Ding W.,
  Gao C., Ge C., Ge W., Guo Z., Huang Q., Huang J., Huang F., Hui B., Jiang S.,
  Li Z., Li M., Li M., Li K., Lin Z., Lin J., Liu X., Liu J., Liu C., Liu Y.,
  Liu D., Liu S., Lu D., Luo R., Lv C., Men R., Meng L., Ren X., Ren X., Song
  S., Sun Y., Tang J., Tu J., Wan J., Wang P., Wang P., Wang Q., Wang Y., Xie
  T., Xu Y., Xu H., Xu J., Yang Z., Yang M., Yang J., Yang A., Yu B., Zhang F.,
  Zhang H., Zhang X., Zheng B., Zhong H., Zhou J., Zhou F., Zhou J., Zhu Y.,
  Zhu K.}:
\newblock {Qwen3-VL} technical report, 2025.
\newblock \href {https://doi.org/10.48550/arXiv.2511.21631}
  {\path{doi:10.48550/arXiv.2511.21631}}.

\bibitem[BEC{\etalchar{*}}18]{Bradley2018Visualization}
\textsc{Bradley A.~J., {El-Assady} M., Coles K., Alexander E., Chen M., Collins
  C., J{\"a}nicke S., Wrisley D.~J.}:
\newblock Visualization and the digital humanities.
\newblock \emph{IEEE Computer Graphics and Applications 38}, 6 (2018), 26--38.
\newblock \href {https://doi.org/10.1109/MCG.2018.2878900}
  {\path{doi:10.1109/MCG.2018.2878900}}.

\bibitem[BHJ09]{Bastian2009Gephi}
\textsc{Bastian M., Heymann S., Jacomy M.}:
\newblock {Gephi}: An open source software for exploring and manipulating
  networks.
\newblock In \emph{Proceedings of the International AAAI Conference on Web and
  Social Media} (Mar. 2009), vol.~3, pp.~361--362.
\newblock \href {https://doi.org/10.1609/icwsm.v3i1.13937}
  {\path{doi:10.1609/icwsm.v3i1.13937}}.

\bibitem[Bur01]{Burke2001Eyewitnessing}
\textsc{Burke P.}:
\newblock \emph{Eyewitnessing: The Uses of Images as Historical Evidence}.
\newblock Cornell University Press, 2001.

\bibitem[CDGP22]{Charles2022Exploring}
\textsc{Charles L., Daudin G., Girard P., Plique G.}:
\newblock Exploring the transformation of french trade in the long eighteenth
  century (1713--1823): The {TOFLIT18} project.
\newblock \emph{Historical Methods: A Journal of Quantitative and
  Interdisciplinary History 55}, 4 (2022), 228--258.
\newblock \href {https://doi.org/10.1080/01615440.2022.2032522}
  {\path{doi:10.1080/01615440.2022.2032522}}.

\bibitem[Cha24]{Charmaz2024Constructing}
\textsc{Charmaz K.}:
\newblock \emph{Constructing Grounded Theory}, 3~ed.
\newblock Introducing Qualitative Methods. SAGE Publications, Thousand Oaks,
  CA, USA, 2024.

\bibitem[{Cla}97]{Clarivate1997Web}
\textsc{{Clarivate}}:
\newblock Web of science core collection.
\newblock
  \url{https://clarivate.com/academia-government/scientific-and-academic-research/research-discovery-and-referencing/web-of-science/web-of-science-core-collection/},
  1997.

\bibitem[CLL{\etalchar{*}}21]{Chen2021VIS30K}
\textsc{Chen J., Ling M., Li R., Isenberg P., Isenberg T., Sedlmair M.,
  M{\"o}ller T., Laramee R.~S., Shen H.-W., Wunsche K., Wang Q.}:
\newblock {VIS30K}: A collection of figures and tables from {IEEE}
  {Visualization} {Conference} publications.
\newblock \emph{IEEE Transactions on Visualization and Computer Graphics 27}, 9
  (2021), 3826--3833.
\newblock \href {https://doi.org/10.1109/TVCG.2021.3054916}
  {\path{doi:10.1109/TVCG.2021.3054916}}.

\bibitem[Col94]{Collingwood1994Idea}
\textsc{Collingwood R.~G.}:
\newblock \emph{The Idea of History: With Lectures 1926-1928}.
\newblock Oxford University Press, 1994.

\bibitem[Cor19]{Correll2019Ethical}
\textsc{Correll M.}:
\newblock Ethical dimensions of visualization research.
\newblock In \emph{Proceedings of the CHI Conference on Human Factors in
  Computing Systems} (2019), ACM, pp.~1--13.
\newblock \href {https://doi.org/10.1145/3290605.3300418}
  {\path{doi:10.1145/3290605.3300418}}.

\bibitem[DFK{\etalchar{*}}26]{Dawson2026Qgis}
\textsc{Dawson N., Fischer J., Kuhn M., Pasotti A., Rouzaud D., {mhugent}, Bruy
  A., Sutton T., Dobias M., Pellerin M., Rouault E., Olaya V., Blottiere P.,
  Macho W., Blazek R., Bartoletti L., Sherman G., {Sant-anna} H., Cabieces J.,
  Woodrow N., {signedav}, {rldhont}, Natsis S., Shaffer L., Felder J., Belgacem
  N., Santilli S., Larosa S., Mani S., Jurgiel B.}:
\newblock {qgis}/{QGIS}: 3.44.9.
\newblock Zenodo, Apr. 2026.
\newblock \href {https://doi.org/10.5281/zenodo.19401570}
  {\path{doi:10.5281/zenodo.19401570}}.

\bibitem[DK16]{DIgnazio2016Feminist}
\textsc{D'Ignazio C., Klein L.~F.}:
\newblock Feminist data visualization.
\newblock In \emph{Proceedings of the IEEE VIS Workshop on Visualization for
  the Digital Humanities} (2016), IEEE.

\bibitem[Dru11]{Drucker2011Humanities}
\textsc{Drucker J.}:
\newblock Humanities approaches to graphical display.
\newblock \emph{Digital Humanities Quarterly 5}, 1 (2011), 1--21.
\newblock \href {https://doi.org/10.63744/r4ysrh7ae534}
  {\path{doi:10.63744/r4ysrh7ae534}}.

\bibitem[DWS{\etalchar{*}}23]{Deng2023VisImages}
\textsc{Deng D., Wu Y., Shu X., Wu J., Fu S., Cui W., Wu Y.}:
\newblock {VisImages}: A fine-grained expert-annotated visualization dataset.
\newblock \emph{IEEE Transactions on Visualization and Computer Graphics 29}, 7
  (2023), 3298--3311.
\newblock \href {https://doi.org/10.1109/TVCG.2022.3155440}
  {\path{doi:10.1109/TVCG.2022.3155440}}.

\bibitem[{Esr}99]{Esri1999ArcMap}
\textsc{{Esri}}:
\newblock {ArcMap}, 1999.

\bibitem[Ewa16]{Ewalt2016Image}
\textsc{Ewalt J.}:
\newblock Image as evidence: A citation analysis of visual resources in
  {American} history scholarship, 2010--2014.
\newblock \emph{Art Documentation: Journal of the Art Libraries Society of
  North America 35}, 2 (2016), 206--217.
\newblock \href {https://doi.org/10.1086/688723} {\path{doi:10.1086/688723}}.

\bibitem[FE95]{Fogel1995Time}
\textsc{Fogel R.~W., Engerman S.~L.}:
\newblock \emph{Time on the Cross: The Economics of American Negro Slavery}.
\newblock W. W. Norton \& Company, 1995.

\bibitem[Fis23]{Fischer2023Network}
\textsc{Fischer L.}:
\newblock \emph{Network Visualization and the Labor of Reference Work: Three
  Case Studies Touching Medieval and Early Modern Book History}.
\newblock PhD thesis, The University of Texas at Austin, 2023.

\bibitem[GAB{\etalchar{*}}22]{Giovannetti2022Ontology}
\textsc{Giovannetti E., Albanesi D., Bellandi A., Dattilo D., Del~Grosso A.~M.,
  Marchi S.}:
\newblock An ontology of masters of the {Babylonian Talmud}.
\newblock \emph{Digital Scholarship in the Humanities 37}, 3 (2022), 725--737.
\newblock \href {https://doi.org/10.1093/llc/fqab043}
  {\path{doi:10.1093/llc/fqab043}}.

\bibitem[Gra22]{Grandjean2022Data}
\textsc{Grandjean M.}:
\newblock Data visualization for history.
\newblock In \emph{Handbook of Digital Public History}, Noiret S., Tebeau M.,
  Zaagsma G., (Eds.). De Gruyter Oldenbourg, 2022, pp.~291--300.
\newblock \href {https://doi.org/10.1515/9783110430295-024}
  {\path{doi:10.1515/9783110430295-024}}.

\bibitem[HFM16]{Hinrichs2016Speculative}
\textsc{Hinrichs U., Forlini S., Moynihan B.}:
\newblock Speculative practices: Utilizing {InfoVis} to explore untapped
  literary collections.
\newblock \emph{IEEE Transactions on Visualization and Computer Graphics 22}, 1
  (Jan. 2016), 429--438.
\newblock \href {https://doi.org/10.1109/TVCG.2015.2467452}
  {\path{doi:10.1109/TVCG.2015.2467452}}.

\bibitem[HFM19]{Hinrichs2019Defense}
\textsc{Hinrichs U., Forlini S., Moynihan B.}:
\newblock In defense of sandcastles: Research thinking through visualization in
  digital humanities.
\newblock \emph{Digital Scholarship in the Humanities 34}, Supplement\_1
  (2019), i80--i99.
\newblock \href {https://doi.org/10.1093/llc/fqy051}
  {\path{doi:10.1093/llc/fqy051}}.

\bibitem[HTLF20]{Hendricks2020Crossref}
\textsc{Hendricks G., Tkaczyk D., Lin J., Feeney P.}:
\newblock {Crossref}: The sustainable source of community-owned scholarly
  metadata.
\newblock \emph{Quantitative Science Studies 1}, 1 (2020), 414--427.
\newblock \href {https://doi.org/10.1162/qss_a_00022}
  {\path{doi:10.1162/qss_a_00022}}.

\bibitem[J{\"a}n16]{Janicke2016Valuable}
\textsc{J{\"a}nicke S.}:
\newblock Valuable research for visualization and digital humanities: A
  balancing act.
\newblock In \emph{Proceedings of the IEEE VIS Workshop on Visualization for
  the Digital Humanities} (2016), IEEE.

\bibitem[JFCS17]{Jaenicke2017Visual}
\textsc{J{\"a}nicke S., Franzini G., Cheema M.~F., Scheuermann G.}:
\newblock Visual text analysis in digital humanities.
\newblock \emph{Computer Graphics Forum 36}, 6 (2017), 226--250.
\newblock \href {https://doi.org/10.1111/cgf.12873}
  {\path{doi:10.1111/cgf.12873}}.

\bibitem[JMJ19]{Jobin2019DocFigure}
\textsc{Jobin K.~V., Mondal A., Jawahar C.~V.}:
\newblock {DocFigure}: A dataset for scientific document figure classification.
\newblock In \emph{Proceedinngs of the International Conference on Document
  Analysis and Recognition Workshops} (2019), vol.~1, IEEE, pp.~74--79.
\newblock \href {https://doi.org/10.1109/ICDARW.2019.00018}
  {\path{doi:10.1109/ICDARW.2019.00018}}.

\bibitem[Kle13]{Klein2013Image}
\textsc{Klein L.~F.}:
\newblock The image of absence: Archival silence, data visualization, and
  {James} {Hemings}.
\newblock \emph{American Literature 85}, 4 (2013), 661--688.
\newblock \href {https://doi.org/10.1215/00029831-2367310}
  {\path{doi:10.1215/00029831-2367310}}.

\bibitem[Kli22]{Kliger2022Translating}
\textsc{Kliger G.}:
\newblock Translating god on the borders of sovereignty.
\newblock \emph{The American Historical Review 127}, 3 (2022), 1102--1130.
\newblock \href {https://doi.org/10.1093/ahr/rhac220}
  {\path{doi:10.1093/ahr/rhac220}}.

\bibitem[KRP{\etalchar{*}}23]{Karishma2023ACLFig}
\textsc{Karishma Z., Rohatgi S., Puranik K.~S., Wu J., Giles C.~L.}:
\newblock {ACL}-{Fig}: A dataset for scientific figure classification.
\newblock In \emph{Proceedings of the AAAI Workshop on Scientific Document
  Understanding} (2023).

\bibitem[LBT{\etalchar{*}}18]{Lamqaddam2018When}
\textsc{Lamqaddam H., Brosens K., Truyen F., Beerens J., {de Prekel} I., Aerts
  J., Verbert K.}:
\newblock When the tech kids are running too fast: Data visualisation through
  the lens of art history research.
\newblock In \emph{Proceedings of the IEEE VIS Workshop on Visualization for
  the Digital Humanities} (2018), IEEE.

\bibitem[Lee07]{Lee2007Boundary}
\textsc{Lee C.~P.}:
\newblock Boundary negotiating artifacts: Unbinding the routine of boundary
  objects and embracing chaos in collaborative work.
\newblock \emph{Computer Supported Cooperative Work 16}, 3 (2007), 307--339.
\newblock \href {https://doi.org/10.1007/s10606-007-9044-5}
  {\path{doi:10.1007/s10606-007-9044-5}}.

\bibitem[LVMVA{\etalchar{*}}21]{Lamqaddam2021Introducing}
\textsc{Lamqaddam H., Vande~Moere A., Vanden~Abeele V., Brosens K., Verbert
  K.}:
\newblock Introducing layers of meaning ({LoM}): A framework to reduce semantic
  distance of visualization in humanistic research.
\newblock \emph{IEEE Transactions on Visualization and Computer Graphics 27}, 2
  (2021), 1084--1094.
\newblock \href {https://doi.org/10.1109/TVCG.2020.3030426}
  {\path{doi:10.1109/TVCG.2020.3030426}}.

\bibitem[Mar95]{Martinez1995Imaging}
\textsc{Martinez K.}:
\newblock Imaging the past: Historians, visual images and the contested
  definition of history.
\newblock \emph{Visual Resources 11}, 1 (1995), 21--45.
\newblock \href {https://doi.org/10.1080/01973762.1995.9658317}
  {\path{doi:10.1080/01973762.1995.9658317}}.

\bibitem[Men11]{Mengel2011Plague}
\textsc{Mengel D.~C.}:
\newblock A plague on {Bohemia}? mapping the black death.
\newblock \emph{Past \& Present 211}, 1 (2011), 3--34.
\newblock \href {https://doi.org/10.1093/pastj/gtq069}
  {\path{doi:10.1093/pastj/gtq069}}.

\bibitem[MSPP25]{Mordechai2025Systematic}
\textsc{Mordechai L., Stahl A., Pyzyk M., Pelle I.~C.}:
\newblock Systematic bias in humanities datasets: Ancient and medieval coin
  finds in the {FLAME} project.
\newblock \emph{Digital Humanities Quarterly 019}, 1 (2025).

\bibitem[Mun14]{Munzner2014Visualization}
\textsc{Munzner T.}:
\newblock \emph{Visualization Analysis and Design}.
\newblock CRC Press, 2014.

\bibitem[MZY{\etalchar{*}}25]{Mei2025ZuantuSet}
\textsc{Mei X., Zhang Y., Yang C., Shi R., Yuan X.}:
\newblock {ZuantuSet}: A collection of historical chinese visualizations and
  illustrations.
\newblock In \emph{Proceedings of the CHI Conference on Human Factors in
  Computing Systems} (2025), ACM.
\newblock \href {https://doi.org/10.1145/3706598.3713276}
  {\path{doi:10.1145/3706598.3713276}}.

\bibitem[PLP{\etalchar{*}}23]{Panagiotidou2023Communicating}
\textsc{Panagiotidou G., Lamqaddam H., Poblome J., Brosens K., Verbert K.,
  Vande~Moere A.}:
\newblock Communicating uncertainty in digital humanities visualization
  research.
\newblock \emph{IEEE Transactions on Visualization and Computer Graphics 29}, 1
  (2023), 635--645.
\newblock \href {https://doi.org/10.1109/TVCG.2022.3209436}
  {\path{doi:10.1109/TVCG.2022.3209436}}.

\bibitem[RG10]{Rosenberg2010Cartographies}
\textsc{Rosenberg D., Grafton A.}:
\newblock \emph{Cartographies of Time: A History of the Timeline}, 1~ed.
\newblock Princeton Architectural Press, 2010.

\bibitem[RKH{\etalchar{*}}21]{Radford2021Learning}
\textsc{Radford A., Kim J.~W., Hallacy C., Ramesh A., Goh G., Agarwal S.,
  Sastry G., Askell A., Mishkin P., Clark J., Krueger G., Sutskever I.}:
\newblock Learning transferable visual models from natural language
  supervision.
\newblock In \emph{Proceedings of the International Conference on Machine
  Learning} (2021), Meila M., Zhang T., (Eds.), vol.~139, PMLR, pp.~8748--8763.

\bibitem[Smi07]{Smith2007Overview}
\textsc{Smith R.}:
\newblock An overview of the {Tesseract} {OCR} engine.
\newblock In \emph{Proceedings of the International Conference on Document
  Analysis and Recognition} (2007), vol.~2, pp.~629--633.
\newblock \href {https://doi.org/10.1109/ICDAR.2007.4376991}
  {\path{doi:10.1109/ICDAR.2007.4376991}}.

\bibitem[SMRB07]{St-Hilaire2007Geocoding}
\textsc{{St-Hilaire} M., Moldofsky B., Richard L., Beaudry M.}:
\newblock Geocoding and mapping historical census data: The geographical
  component of the {Canadian} century research infrastructure.
\newblock \emph{Historical Methods: A Journal of Quantitative and
  Interdisciplinary History 40}, 2 (2007), 76--91.
\newblock \href {https://doi.org/10.3200/HMTS.40.2.76-91}
  {\path{doi:10.3200/HMTS.40.2.76-91}}.

\bibitem[Sta10]{Star2010This}
\textsc{Star S.~L.}:
\newblock This is not a boundary object: Reflections on the origin of a
  concept.
\newblock \emph{Science, Technology, \& Human Values 35}, 5 (2010), 601--617.
\newblock \href {https://doi.org/10.1177/0162243910377624}
  {\path{doi:10.1177/0162243910377624}}.

\bibitem[Sto19]{Storey2019Cartographically}
\textsc{Storey E.~A.}:
\newblock Cartographically reconstructing surveys of community land grants in
  {New Mexico} to support historical research and political discourse.
\newblock \emph{Historical Methods: A Journal of Quantitative and
  Interdisciplinary History 52}, 2 (2019), 95--109.
\newblock \href {https://doi.org/10.1080/01615440.2018.1502641}
  {\path{doi:10.1080/01615440.2018.1502641}}.

\bibitem[{The}19]{TheAmericanHistoricalReview2019Index}
\textsc{{The American Historical Review}}:
\newblock Index of topics {October} 2019.
\newblock \emph{The American Historical Review 124}, 4 (2019), 1587--1589.
\newblock \href {https://doi.org/10.1093/ahr/rhz1009}
  {\path{doi:10.1093/ahr/rhz1009}}.

\bibitem[THSN22]{Toth2022Studying}
\textsc{T{\'o}th G.~M., Hempel T., Somandepalli K., Narayanan S.}:
\newblock Studying large-scale behavioral differences in {Auschwitz-Birkenau}
  with simulation of gendered narratives.
\newblock \emph{Digital Humanities Quarterly 016}, 3 (2022).
\newblock \href {https://doi.org/10.63744/53wts95vs45u}
  {\path{doi:10.63744/53wts95vs45u}}.

\bibitem[TW20]{Tsui2020Harvesting}
\textsc{Tsui L.~H., Wang H.}:
\newblock Harvesting big biographical data for {Chinese} history: The {China
  Biographical Database (CBDB)}.
\newblock \emph{Journal of Chinese History 4}, 2 (2020), 505--511.
\newblock \href {https://doi.org/10.1017/jch.2020.21}
  {\path{doi:10.1017/jch.2020.21}}.

\bibitem[ZCZ{\etalchar{*}}25]{Zhang2025VisTaxa}
\textsc{Zhang Y., Chen X., Zheng W., Guo Y., Li G., Chen S., Yuan X.}:
\newblock {VisTaxa}: Developing a taxonomy of historical visualizations.
\newblock \emph{IEEE Transactions on Visualization and Computer Graphics 31}, 6
  (2025), 3850--3862.
\newblock \href {https://doi.org/10.1109/TVCG.2025.3567132}
  {\path{doi:10.1109/TVCG.2025.3567132}}.

\bibitem[ZJX{\etalchar{*}}24]{Zhang2024OldVisOnline}
\textsc{Zhang Y., Jiang R., Xie L., Zhao Y., Liu C., Ding T., Chen S., Yuan
  X.}:
\newblock {OldVisOnline}: Curating a dataset of historical visualizations.
\newblock \emph{IEEE Transactions on Visualization and Computer Graphics 30}, 1
  (2024), 551--561.
\newblock \href {https://doi.org/10.1109/TVCG.2023.3326908}
  {\path{doi:10.1109/TVCG.2023.3326908}}.

\bibitem[ZTJY19]{Zhong2019PubLayNet}
\textsc{Zhong X., Tang J., Jimeno~Yepes A.}:
\newblock {PubLayNet}: Largest dataset ever for document layout analysis.
\newblock In \emph{Proceedings of the International Conference on Document
  Analysis and Recognition} (2019), IEEE, pp.~1015--1022.
\newblock \href {https://doi.org/10.1109/ICDAR.2019.00166}
  {\path{doi:10.1109/ICDAR.2019.00166}}.

\end{thebibliography}
\fi
\fi

\end{document}